\pdfoutput=1
\documentclass[twocolumn,apj]{openjournal}

\let\oldthebibliography\thebibliography
\renewcommand{\thebibliography}[1]{%
  \oldthebibliography{#1}
  \let\oldbibitem\bibitem
  \let\oldtextsc\textsc
  \def\oldbbland{et}
  \newcounter{authorcount}
  \def\bibitem[##1]##2{%
    \let\textsc\oldtextsc
    \let\bbland\oldbbland
    \oldbibitem[##1]{##2}%
    \let\textsc\mytextsc%
    \let\bbland\mybbland
    \setcounter{authorcount}{0}
  }
  \def\mybbland{\setcounter{authorcount}{0}\oldbbland}
  \def\dropetal##1.{ \bbletal}
  \def\mytextsc##1{%
    \oldtextsc{##1}%
    \stepcounter{authorcount}%
    \ifnum\value{authorcount}=5\relax%
      \expandafter\dropetal%
    \fi%
  }%
}
\usepackage{lipsum}

\usepackage{xcolor}
\usepackage{textgreek}
\usepackage[utf8]{inputenc}
\usepackage[english]{babel}
\usepackage{amsmath}

\usepackage{hyperref}
\hypersetup{
    unicode, 
    colorlinks=true,
    linkcolor=linkcolor,
    citecolor=linkcolor,
    filecolor=linkcolor,
    urlcolor=linkcolor,
}

\usepackage{color,colortbl}
\definecolor{linkcolor}{rgb}{0.0,0.3,0.5}
\usepackage{tensind}
\tensordelimiter{?}
\DeclareGraphicsExtensions{.bmp,.png,.jpg,.pdf}
\usepackage{verbatim}
\usepackage[normalem]{ulem}
\usepackage{orcidlink}
\usepackage{soul}

\graphicspath{ {./figs/} }
\setcitestyle{notesep={ }}
\begin{document}
\title{Augmenting Galaxy Catalogues using Machine Learning in the FLAMINGO Simulations}

\author{Elliot Scott$^{1}$* \orcidlink{0009-0006-4813-590X}, Marika Asgari$^{1}$\orcidlink{0000-0002-3064-083X}, Joop Schaye$^{2}$\orcidlink{0000-0002-0668-5560}, Matthieu Schaller$^{2,3}$\orcidlink{0000-0002-2395-4902}, Markus Michael Rau$^{1}$\orcidlink{0000-0003-3709-1324}}
\affiliation{
$^1$School of Mathematics, Statistics and Physics, Newcastle University, Herschel Building, NE1 7RU, Newcastle-upon-Tyne, UK\\
$^2$Leiden Observatory, Leiden University, P.O.Box 9513, 2300RA Leiden, The Netherlands\\
$^3$Lorentz Institute for Theoretical Physics, Leiden University, PO Box 9506, NL-2300 RA Leiden, The Netherlands
}

\altaffiltext{*}{e.scott18@newcastle.ac.uk}

\begin{abstract}
Hydrodynamic simulations are instrumental in the analysis of observations such as galaxy clustering used for the estimation of cosmological parameters. However, for these purposes, the simulations need both a large box size and high resolution, which is computationally prohibitive. In this work we use the connection between galaxies and their host dark matter halos to populate dark-matter-only simulations, using data from the FLAMINGO simulations suite. We employ the Extremely Randomized Trees algorithm to model the connection between the mass profile, history and environment of dark matter halos and the probability distribution of the stellar mass of the galaxy contained in the halo. We use halos matched between a dark-matter-only simulation and the corresponding hydrodynamical simulation to train a model for centrals. For satellites, we employ a Gaussian Processes emulator that simultaneously optimises the hyperparameters of a model and selects for the dark matter features that are predictive of the stellar mass while being insensitive to the differences between the hydrodynamic and dark-matter-only simulations, then train a model on only these features. We then apply these models to produce a new galaxy catalogue for the 2.8\;Gpc FLAMINGO-10k dark-matter-only simulation as a product of this work. We demonstrate that this catalogue reproduces the statistics of the galaxy-halo connection predicted by the smaller volume hydrodynamical simulation on which the model was trained, including the stellar mass function, stellar mass-dark matter mass relation, 2-point correlation functions and correlations between stellar mass and dark matter concentration, half mass radius and dark matter mass within 3\;Mpc.
\end{abstract}


\maketitle

\section{Introduction}
\label{sec:intro}

Observations of large-scale structure in the universe indicate that the structure is dominated by cold dark matter. However, these observations rely on the luminous tracers of the inferred underlying dark matter structure. A number of probes for large scale structure have been employed in past works: for larger-scale clustering, cosmic shear from galaxy surveys \citep{wrightKiDSLegacyCosmologicalConstraints2025f, collaborationDarkEnergySurvey2026, dalalHyperSuprimeCamYear2023} has been a common probe, whereas galaxy clustering \citep{adameDESI2024FullShape2025, alamClusteringGalaxiesCompleted2017} and cluster counts \citep{adePlanck2015Results2016, artisSRGEROSITAAllSky2025} are used for studying more localised clustering. In modern cosmological analysis, a standard combined probe of large-scale structure is $3\times2$ point statistics, which is the combination of galaxy clustering, galaxy-galaxy lensing and cosmic shear \citep[][]{moreHyperSuprimeCamYear2023, descollaborationDarkEnergySurvey2026}.
In this work, we focus on galaxy statistics, which mainly affects cluster counts, galaxy clustering and galaxy-galaxy lensing (with a small effect on cosmic shear also due to intrinsic alignment).

It is necessary for modelling of $3\times2$ point statistics to have knowledge of the matter and galaxy distributions over a wide range of scales. The Halo Model \citep[see][for a review]{asgariHaloModelCosmology2023} can do so, but requires calibration to simulations or data. In the default Halo Model, it is assumed that all dark matter is contained in spherically-symmetric halos, although this assumption can be relaxed moving away from the default model. Galaxies populate these halos with a central galaxy surrounded by a number of satellites; a standard model of the occupation statistics of these galaxies is the five-parameter model in \cite{zhengTheoreticalModelsHalo2005}. When calibrating the Halo Model using simulations, these simulations must be hydrodynamic due to the need for knowledge of galaxy distributions, which comes with a large computational cost. Furthermore, hydrodynamic simulations with greater resolution are capable of resolving galaxies over a greater range of masses, and simulations with a larger box size have a larger sample of halos, which can be used to calibrate a halo model more precisely. Both of these aspects of hydrodynamic simulations also increase the computational cost. 

Since it is computationally cheaper to run a dark matter only (DMO) simulation instead of a hydrodynamic simulation, it is desirable to have a mechanism to inject baryonic information into a DMO simulation. Studying the connection between the properties of dark matter subhalos and the galaxies that reside within them in a hydrodynamic simulation makes it possible to create a predictive pipeline which can ingest the properties of the subhalos in a DMO simulation and estimate baryonic properties of the galaxies within those subhalos. This connection is typically referred to as the galaxy-halo connection.

Inferring the galaxy-halo connection from observation is non-trivial but several past works have attempted to do so. \cite{yangGalaxyGroupsSDSS2008} use data from the Sloan Digital Sky Survey \citep[SDSS,][]{yorkSloanDigitalSky2000} Data Release 4 \citep{FourthDataRelease2006} to constrain the Conditional Luminosity Function (CLF, the luminosity distribution of galaxies at a given halo mass). \cite{cacciatoGalaxyClusteringGalaxy2009} use an analytic model of the CLF to predict galaxy-galaxy lensing signals, which are found to match SDSS observations. \cite{mosterGalacticStarFormation2013} characterise the relation between dark matter mass and galaxy mass in halos with both simulations and observational data. They use the Millennium Simulation \citep{springelSimulatingJointEvolution2005} for massive halos and the Millennium-II Simulation \citep{boylan-kolchinResolvingCosmicStructure2009} for low-mass halos. They also use SMF measurements from \cite{liDistributionStellarMass2009}, \cite{perez-gonzalezStellarMassAssembly2008} and \cite{santiniEvolvingSlopeStellar2012}.

There have been several methods employed to model the galaxy-halo connection in previous works, such as Subhalo Abundance Matching, Semi-Analytic Models and machine learning. Subhalo Abundance Matching  \citep[SHAM, first demonstrated in works such as][]{valeLinkingHaloMass2004, shankarNewRelationshipsGalaxy2006, conroyModelingLuminositydependentGalaxy2006} makes the assumption that there exists a one-to-one and monotonic mapping between some subhalo property and some baryonic property, along with an intrinsic scatter. Two functions can then be fit which model the mean and standard deviation of the baryonic property at a given value of the dark matter property. This then gives the parameters of a normal distribution which can be sampled from. The simplest case of this would be the subhalo mass and the stellar mass, although typically a subhalo property such as the maximum of the circular velocity profile ($V_{\text{max}}$) is preferred \citep[e.g.][]{chaves-monteroSubhaloAbundanceMatching2016}. Often, the maximum of the circular velocity profile over all time ($V_{\text{peak}}$) or at times fulfilling a certain condition are used. In \cite{chaves-monteroSubhaloAbundanceMatching2016}, variants of the timeframe to take the maximum of the circular velocity were compared using a simulation from the EAGLE suite \citep{schayeEAGLEProjectSimulating2015a}. Subhalo Abundance Matching gives some physical intuition in how the dark matter property relates to the baryonic property through the two functions that are fitted, and is computationally cheap and simple to implement. \cite{reddickCONNECTIONGALAXIESDARK2013} found that SHAM is able to reproduce observations from \cite{abazajianSeventhDataRelease2009a}, with the best model using $V_{\text{peak}}$ as the dark matter parameter. \cite{ortega-martinezInvestigatingGalaxyhaloConnection2025} used an extension of SHAM described in \cite{ortega-martinezSHAMeSFPredictingClustering2024} to reproduce the clustering of DESI Emission Line Galaxies \citep{2024desi}. \cite{boreikoTestingSubhaloAbundance2026} used a SHAM model with a parametrised halo response to galaxy formation to estimate the likelihood of the observed maximum of the circular velocity for the 175 galaxies in the SPARC dataset \citep{lelliSPARCMassModels2016}. They found the SHAM scatter to be in tension with low-mass clustering constraints even when accounting for selection effects.

For small expansions to the input parameter space, it is possible to construct a new parameter which is a function of two or more subhalo properties \citep[e.g. in][]{chuangNewStellarMass2023}. \cite{chaves-monteroSubhaloAbundanceMatching2016} remarks on the dependence on halo mass for the assumed relation between the maximum of the subhalo circular velocity and the galactic stellar mass. However, manually constructing a sensible relation between multiple subhalo properties is not realistic for a larger set of input parameters, necessitating more sophisticated techniques. Furthermore, we can more accurately model the galaxy-halo connection if we can back off from the assumption of a one-to-one and monotonic mapping between dark matter and baryonic properties.

Several groups have worked to use machine learning techniques to model the connection between subhalos and galaxies. For example, \cite{jagvaralGalaxiesHaloesGraph2022} used a combination of a Mixture Density Network and a Generative Adversarial Network to estimate galactic axis ratios and the direction of the main axis (among other features) trained on TNG-100 \citep[TNG-100 is detailed in ][]{pillepichFirstResultsIllustrisTNG2018}.\cite{lovellMachineLearningApproach2022} and \cite{icaza-lizaolaSparseRegressionApproach2023} both trained on EAGLE \citep{schayeEAGLEProjectSimulating2015a} using a random forest and sparse regression model respectively to estimate stellar mass \citep[as well as other galactic properties
in ][]{lovellMachineLearningApproach2022}. In principle, these techniques are all similar to SHAM, though the physical relations are more obscured. Much work has gone into coming up with new sets of parameters to input into these models, aiming to capture historic and present information about the dark matter subhalo where the desired galaxy resides. Much of this work has been done with older simulations, often trained with smaller box sizes - considering both TNG-100 used for \cite{jagvaralGalaxiesHaloesGraph2022} and the EAGLE runs used for \cite{lovellMachineLearningApproach2022} and \cite{icaza-lizaolaSparseRegressionApproach2023} have box side length of $75\space\text{Mpc}\;h^{-1}$ and $100\space\text{Mpc}\;h^{-1}$ respectively.

The machine learning model architectures used in previous works have varied. Several groups have used variants of Random Forest models \citep{hoRandomDecisionForests1995}. A Random Forest is an ensemble of decision trees that attempt to sort samples based on split points in the data. For classification tasks, the model can select the most frequently estimated class from the trees. For regression tasks, the model can average the outputs of all the trees. In \cite{lovellMachineLearningApproach2022}, the variant Extremely Randomized Trees \citep{geurtsExtremelyRandomizedTrees2006} is used. In a conventional Random Forest the choice of features to use in a given tree is random, but the split points are deterministic. In Extremely Randomized Trees, the feature choice and split points are both randomised. \cite{lovellMachineLearningApproach2022} compared their random forest model with two isotonic regression models, with the two models using the two most common SHAM inputs, subhalo mass and the maximum of the circular velocity profile. For all 6 galaxy properties tested, the random forest showed a higher Pearson correlation coefficient between the predicted galaxy property values and the true values than both isotonic regression models.

Many of these aforementioned works predict the value of the desired galaxy property. However, there is an intrinsic scatter in the property. This is captured in SHAM as the scatter in the baryonic property and is estimated as a function of the dark matter property. However, this is less trivial when using several dark matter properties. In \cite{jagvaralGalaxiesHaloesGraph2022}, a combination of a Mixture Density Network and a Generative Adverserial Network is used to estimate the probability distribution function of baryonic properties. In that work and the subsequent work \citep{jagvaralGeometricDeepLearning2025}, they construct a graph to model for use in a Graph Neural Network. Alternatively, \cite{rodriguesHighfidelityReproductionCentral2023} elects to make an estimate on the probability distribution of the baryonic properties by treating the problem as a classification task. In this case, the model assigns a probability for each class, and the combination of all these probabilities gives an estimator for the probability distribution. A regressor model will typically be optimised to minimise the mean squared error, leading to the model estimating the mean of the probability distribution of the baryonic property. This reduces the scatter in that property for given dark matter property values e.g. the scatter in the conditional stellar mass function would be underestimated by a model that has been optimised to minimise the mean squared error. Instead, estimating the probability distribution avoids this issue. \cite{rodriguesExploringHalogalaxyConnection2025} build on the previous work by defining the categorical distribution using a Voronoi cell-based hierarchical scheme. \cite{stiskalekScatterGalaxyHalo2022} use an ensemble of neural networks to also make predictions for the scatter in the baryonic properties of interest. They find that a neural network trained only on the dark matter mass of halos is able to accurately recover the scatter in the stellar mass-halo mass relation. However, the scatter is no longer reproduced when training with other dark matter features included, especially at low masses and for satellites.

It should however be noted that cosmological simulations contain an irreducible scatter in scaling relations which is caused by the inherent stochasticisty of the subgrid models, as demonstrated in \cite{borrowIrreducibleScatter}. With large samples of galaxies, they found the median of scaling relations is consistently produced. However, individual galaxy properties such as stellar mass could vary by up to 25\%. Furthermore, \cite{kellerChaosVarianceGalaxy2019} found that small perturbations due to rounding and stochastic effects can produce substantial variation in the distribution of stellar mass. Characterising this scatter is beyond the scope of this work but should be considered when interpreting the results presented.

In this work, we train Extremely Randomized Trees models on the FLAMINGO simulation suite \citep{schayeFLAMINGOProjectCosmological2023, kugelFLAMINGOCalibratingLarge2023, Helly_2026} to predict the probability distribution of stellar masses for halos based on their dark matter properties. The FLAMINGO simulations are described in Section \ref{sec:flamingo}. In Section \ref{sec:train}, we discuss the data processing and hyperparamater optimisation necessary for training our models and optimising their performance when applied to a DMO simulation. In Section \ref{sec:results}, we evaluate the models' performance using FLAMINGO DMO simulations and present new galaxy catalogues. In particular, we produce catalogues using the FLAMINGO-10k DMO simulation run, which has the highest resolution of all FLAMINGO runs and is the largest box size in FLAMINGO at that resolution.\newline\newline\newline

\begingroup 
    \setlength{\tabcolsep}{10pt} 
    \renewcommand{\arraystretch}{1.5} 
    \setlength\extrarowheight{2pt}
\begin{table*}[!t]
\centering
\begin{tabular}{|l|l|l|l|l|l|l|}
\hline
               Run Name & $m_{\rm{g}}/M_{\odot}$ & $m_{\text{CDM}}/M_{\odot}$ & L/cGpc & Hydro/DMO & Cosmology & Feedback Model\\ \hline\hline
L1\_m8         & $1.34 \times 10^8$                     & $7.06 \times 10^8$                           & 1                    & Hydrodynamic   & D3A  & Thermal             \\ \hline
L1\_m8\_DMO    & n/a                                    & $8.40 \times 10^8$                           & 1                    & Dark Matter Only  & D3A  & n/a            \\ \hline
FLAMINGO-10k  & n/a                                    & $8.40 \times 10^8$                           & 2.8                  & Dark Matter Only  & D3A  & n/a            \\ \hline
\end{tabular}
\caption{Table of FLAMINGO runs with the fiducial cosmology and feedback model that are relevant to this work. $m_g$ is the initial baryonic particle mass, $m_{\text{CDM}}$ is the initial dark matter particle mass, L is the comoving box side length. The final three columns indicate whether the simulation was hydrodynamic or dark matter only, its cosmological parameters and its feedback model respectively.}

\label{table:flamingoruns}
\end{table*}
\endgroup

\section{FLAMINGO Simulations}
\label{sec:flamingo}

The FLAMINGO simulations \citep[][]{schayeFLAMINGOProjectCosmological2023} are a leading suite of large-volume hydrodynamic simulations with three particle resolutions and a range of cosmologies and feedback models. All the FLAMINGO data have been publicly released, as described in \cite{Helly_2026}. The hydrodynamic runs contain dark matter, stellar, gas, black hole and neutrino particles which are evolved in time using the \texttt{SWIFT} hydrodynamics solver \citep{schallerSWIFTModernHighlyparallel2024}. For all hydrodynamic runs there are equivalent dark-matter-only (DMO) runs, generated with the same initial conditions and seed, allowing like-for-like comparison between the hydrodynamic and DMO runs.

The simulations have their subgrid parameters calibrated to observations, specifically to the galaxy stellar mass function (SMF) at $z=0$ and the cluster gas mass fraction at $z\approx0.3$. This calibration was done with a Gaussian Processes emulator in \cite{kugelFLAMINGOCalibratingLarge2023}. For the SMF, the calibration lower limit is a stellar mass of $10^{8.67}\rm{M_{\odot}}$, $10^{9.92}\rm{M_{\odot}}$ and $10^{11.17}\rm{M_{\odot}}$ at high, medium and low resolution respectively. The upper stellar mass limit for calibration is $10^{11.5}\rm{M_{\odot}}$ for all resolutions. All resolutions show good agreement with the observational data, with the highest resolution run having the weakest agreement at around 0.2\;dex difference in number density. For the gas mass fraction, the lower mass limit of calibration for all resolutions was $M_{\text{500c}}=10^{13.5}\rm{M_{\odot}}$ (where $M_{\text{500c}}$ is the total cluster mass within a sphere with density 500 times the critical density). The higher mass limits are $10^{13.73}\rm{M_{\odot}}$, $10^{14.36}\rm{M_{\odot}}$, $10^{14.53}\rm{M_{\odot}}$ for the high, medium and low-resolution runs respectively due to the reduced box size used for calibration. Despite this, the high resolution run shows a very similar trend to the other resolutions outside of the mass range for which it was calibrated to. 

Beyond these properties, FLAMINGO reproduces a number of other properties not calibrated for. These include the most massive black hole mass as a function of stellar mass within galaxies, star formation rate density as a function of redshift and specific star formation rate as a function of stellar mass. There are some discrepancies with observational data, such as galaxy half-mass radii being generally overestimated and an overabundance of passive galaxies at low mass.

Within the FLAMINGO suite, there are variations to the AGN feedback models and cosmological parameters. The fiducial feedback model is a thermal model with four subgrid parameters, with a jet model as an alternative. All simulation runs follow the flat $\Lambda\rm{CDM}$ model and the fiducial cosmological model uses the maximum posterior likelihood cosmological parameters from the Dark Energy Survey Year 3 results \citep[][referred to as 'D3A']{abbottDarkEnergySurvey2022}. There are two alternative sets of cosmological parameters, one from \cite{planckcollaborationPlanck2018Results2020a} referred to as 'Planck' and one from \cite{amonConsistentLensingClustering2023} referred to as 'LS8'. The LS8 run notably has a substantially lower value of $S_8$ compared to D3A or Planck.

The details of all relevant FLAMINGO simulations are given in Table \ref{table:flamingoruns}. The simulations used to train and validate our models are L1\_m8 and L1\_m8\_DMO, the highest resolution hydrodynamic simulation and its DMO counterpart respectively. We also make use of the FLAMINGO-10k simulation which has the same resolution as L1\_m8\_DMO but a larger box size.

The data products are stored in snapshots which are separated in redshift by $\Delta z = 0.05$ in the range $z=0-3$, $\Delta z = 0.25$ for $z=3-5$, as well as a small number of snapshots at higher redshift than $z=5$. Halos are first identified at an early snapshot in the simulations using a 3D Friends-of-Friends algorithm. These halos and their mergers are then tracked in time through the subsequent snapshots. Subhalos were identified from the particle data using \texttt{HBT-HERONS} \citep{forouharmorenoAssessingSubhaloFinders2025}. This is a variant of the \texttt{HBT+} algorithm \citep{hanHBTImprovedCode2018} with improved tracking of subhalos. The properties of these subhalos were then computed using the \texttt{SOAP} package \citep[Spherical Overdensity \& Aperture Processing, ][]{mcgibbonSOAPPythonPackage2025}. This gives rise to galaxy and subhalo catalogues which are the principal data input for this work. The \texttt{SOAP} galaxy catalogues contain a large number of properties (mass, maximum of the circular velocity profile, etc.) computed both for all matter and for dark matter or baryonic matter. Since \texttt{HBT-Herons} tracks the subhalos across snapshots, the catalogues include properties involving the history of the subhalo such as the snapshot index of formation and $V_{\text{peak}}$, the maximum of $V_{\text{max}}$ over all time. An additional script has been used to match halos between hydrodynamic and dark-matter-only simulation runs by finding halos with the same particle IDs. As part of this work we make use of the matching between L1\_m8 and L1\_m8\_DMO, though it should be noted that this matching process is only able to match central halos, not satellite subhalos.

The properties of the halo catalogues used in this work are detailed in Table \ref{table:features}.

\begingroup 
    \setlength{\tabcolsep}{10pt} 
    \renewcommand{\arraystretch}{1.5} 
    \setlength\extrarowheight{2pt}
\begin{table*}[!t]
\centering
\begin{tabular}{|l|l|}
\hline
Feature Name & Feature Description \\ \hline\hline
$M_{\rm{DM}}$ & Bound dark matter mass within the subhalo \\ \hline
$V_{\rm{DM,max}}$ & Maximum of the dark matter circular velocity profile for the subhalo at the present snapshot\\ \hline
$V_{\rm{DM,peak}}$ & Maximum of the dark matter circular velocity profile for the subhalo over all past and present snapshots\\ \hline
$z_{\rm{DM,peak}}$ & Redshift of the snapshot where $V_{\rm{DM,peak}}$ occurs\\ \hline
$R_{\rm{DM,max}}$ & Radius at which the dark matter circular velocity is $V_{\rm{DM,max}}$\\ \hline
$M_{\text{DM,max}}$ & Maximum of the bound dark matter mass of the subhalo over all past and present snapshots \\ \hline
$z_{\text{DM,max}}$ & Redshift of the snapshot where $M_{\text{DM,max}}$ occurs \\ \hline
$R_{\rm{DM,1/2}}$ & Radius of a sphere containing half the bound dark matter mass of the subhalo \\ \hline
$M_{\rm{DM,10kpc}}$ & Dark matter mass within 10\;kpc of the halo \\ \hline
$M_{\rm{DM,30kpc}}$ & Dark matter mass within 30\;kpc of the halo \\ \hline
$M_{\rm{DM,50kpc}}$ & Dark matter mass within 50\;kpc of the halo \\ \hline
$M_{\rm{DM,100kpc}}$ & Dark matter mass within 100\;kpc of the halo \\ \hline
$M_{\rm{DM,300kpc}}$ & Dark matter mass within 300\;kpc of the halo \\ \hline
$M_{\rm{DM,500kpc}}$ & Dark matter mass within 500\;kpc of the halo \\ \hline
$M_{\rm{DM,1000kpc}}$ & Dark matter mass within 1000\;kpc of the halo \\ \hline
$M_{\rm{DM,3000kpc}}$ & Dark matter mass within 3000\;kpc of the halo \\ \hline
$z_{\rm{birth}}$ & Redshift of the birth of the subhalo \\ \hline
$V_{\rm{DM,COM}}$ & Velocity of the centre of mass of the bound dark matter in the subhalo \\ \hline
$L_{\rm{DM}}$ & Magnitude of the angular momentum vector of the bound dark matter in the subhalo \\ \hline
$\sigma_{\rm{DM}}$ & 1D velocity dispersion of the bound dark matter in the subhalo \\ \hline
$c_{\rm DM,200m,cent}$ & Dark matter concentration for spherical overdensity with density 200 times the mean density\\ \hline
$c_{\rm DM,200c,cent}$ & Dark matter concentration for spherical overdensity with density 200 times the critical density\\ \hline
$M_{\rm{DM,host, sat}}$ & Dark matter mass of the host central halo \\ \hline
$R_{\rm{DM,ratio,sat}}$ & Ratio of the distance of the satellite from its host to the half-mass radius of the host \\ \hline
$V_{\rm{DM,ratio,sat}}$ & Ratio of the $V_{\rm{DM,peak}}$ for the satellite to $V_{\rm{DM,peak}}$ of the host central \\ \hline
$M_{\star}$ & Bound stellar mass within the halo \\ \hline
\end{tabular}
\caption{The features which are used as inputs and the output of the models. Features which are computed only for centrals are denoted with the subscript 'cent' and features computed only for satellites are denoted with the subscript 'sat'.}
\label{table:features}
\end{table*}
\endgroup

    \begin{figure}[!t]
        \centering
\includegraphics[scale=1]{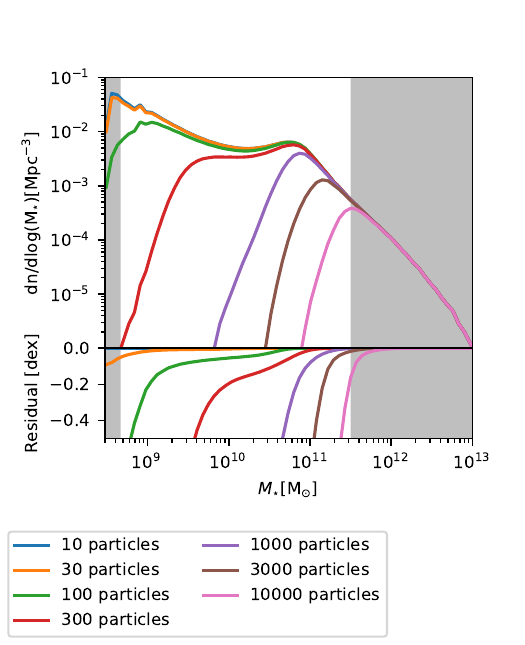}
    	\caption{The galaxy stellar mass function for L1\_m8 shown for a range of minimum dark matter particle thresholds. The bottom panel shows the residual of the SMF compared with that for a particle threshold of 1. The dark grey regions correspond to regions outside which the L1\_m8 was calibrated. Lines for 1 and 3 particles are not shown as they are identical to that for 10 particles. We can see that the lower stellar mass limit for which the stellar mass function is reasonable increases with the particle threshold.}
        \label{fig:smfvspart}
    \end{figure}

\begin{figure*}[!t]
    \centering
    \includegraphics[scale=1]{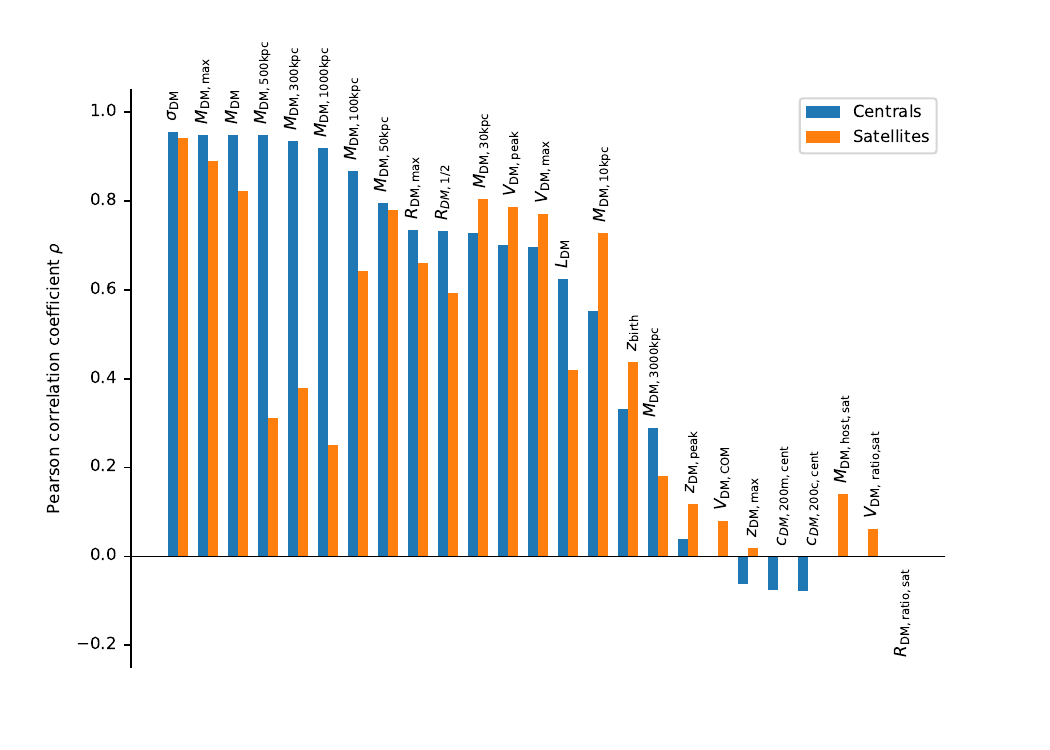}
    \caption{The pearson correlation coefficient (denoted by $\rho$) between all input features and the stellar mass for L1-m8 with a minimum dark matter particle number of 100 imposed. This has been computed separately for centrals (blue) and satellites (orange). The features have been ordered from highest to lowest correlation for centrals. It should be noted that the concentration parameters are defined only for centrals and the final three features are defined only for satellites.}
    \label{fig:corrplot}
\end{figure*}
\section{Optimising and Training the Models}
\label{sec:train}

\subsection{Data Exploration and Preprocessing}
\label{sec:dataex}

The data examined in this section are from the L1\_m8 simulation, which has the highest resolution of any of the FLAMINGO hydrodynamic runs. As described in Table \ref{table:flamingoruns}, this simulation has a comoving box side length of 1 Gpc, initial dark matter particle mass of $7.06\times10^8\rm{M_{\odot}}$ and initial baryonic particle mass of $1.34\times10^8\rm{M_{\odot}}$. This run contains galaxies with the largest range of masses, as well as the largest particle count for a given mass of galaxy. We also use data from L1\_m8\_DMO, which is the DMO counterpart to L1\_m8. It has the same box size, initial conditions and seed as L1\_m8. The dark matter particle mass is $8.4\times10^8\rm{M_{\odot}}$ to maintain the same total mass for the same number of dark matter particles. The majority of central halos in L1\_m8 have been matched to their equivalent halos in L1\_m8\_DMO, although no satellites have been matched between the two simulations. 

The subhalo properties used in this work are detailed in Table \ref{table:features}. Several of these do not exist within the SOAP galaxy catalogues so were computed, including those such as $V_{\text{DM,peak}}$ which required ingesting all previous snapshots as well as the $z=0$ snapshot which the model is trained and tested on. 

We discard all halos from the datasets with fewer than 100 dark matter particles as many SOAP properties are not computed for fewer than 100 particles. Applying this particle threshold also conveniently removes a large proportion of halos which are close to the resolution limit of the simulation, although this also impacts the low stellar mass region of our sample. In Figure \ref{fig:smfvspart} we show how the SMF varies as we vary this minimum particle threshold. At 100 particles, we see less than $0.1\;\rm{dex}$ loss of galaxies above $\sim2\times10^{9}M_{\odot}$.

In Figure \ref{fig:corrplot}, we show the correlations of our input features and the target feature (stellar mass) in L1\_m8, separated for centrals and satellites. The features are ordered by their correlation to the stellar mass for centrals which gives an indication of how predictive these features may be. We note that measures of the dark matter mass have the highest correlations as well as $\sigma_{\rm{DM}}$, $V_{\text{max}}$, $R_{\rm{DM, max}}$ and $V_{\text{DM,peak}}$. This is not surprising as properties such as these are common choices for the dark matter property in SHAM, e.g. in \cite{chaves-monteroSubhaloAbundanceMatching2016}. Measures of the local dark matter mass such as $M_{\rm{DM}, 100kpc}$ have a markedly higher correlation coefficient for centrals compared with satellites, although the reverse is true for $M_{\rm{DM}, 30kpc}$ and especially $M_{\rm{DM}, 10kpc}$.

Given the large number of input features, Principal Component Analysis (PCA) is an obvious approach to reduce their number with minimal information loss. However, we find that there is a substantial drop in the cumulative variance when using small numbers of principal components. Therefore, in order to get minimal information loss, we would need to use a number of principal components similar to the number of input features. Further details are given in Appendix \ref{ap:pca}. 

For all of our models, we take the logarithm of the stellar mass to be the target feature. This was done to ensure that, when we bin by stellar mass, the bins evenly cover all of the orders of magnitude of the stellar mass that we are interested in. Finally, in L1\_m8 we split the data 75\%:25\% into training and test data and then furthermore separate 10\% of the training set (7.5\% of the total data) to be a validation data set. In L1\_m8\_DMO, we separate 10\% of the data to use as the validation data set for hyperparameter optimisation.

\begingroup 
    \setlength{\tabcolsep}{10pt} 
    \renewcommand{\arraystretch}{1.5} 
    \setlength\extrarowheight{2pt}
\begin{table*}[]
\centering
\begin{tabular}{|l|l|l|l|}
\hline
Hyperparameter & Full Range Explored & Central Optimal Value & Satellite Optimal Value \\ \hline\hline
Number of Trees & 10-5000 & 775 & 4223  \\ \hline
Number of Features per Tree & 1-22 (Central) / 1-23 (Satellite) & 14 & 19 \\ \hline
Maximum Depth of Trees & 5-50 & 46 & 72 \\ \hline
Minimum Samples to Split  &2-1000 & 638 & 602 \\ \hline
Number of Stellar Mass Bins &10-500 & 222 & 770 \\ \hline
Data Fraction & $10^{-5}$-$10^{0}$ (sampled logarithmically) &  0.505 & 0.512 \\ \hline
Features used in model & N/A (Central) / All 23 features (Satellite) & N/A & $M_{\rm{DM}}$, $M_{\rm{DM, 1000kpc}}$,  \\&&&$M_{\rm{DM, 3000kpc}}$,  $M_{\rm{DM, max}}$,\\&&&  $V_{\rm{DM,COM}}$,  $M_{\rm{DM,host,sat}}$ \\ \hline
\end{tabular}
\caption{The ranges of hyperparameters explored and the optimal values found, shown for both the centrals and satellites models in \texttt{DMO-opt}. This also includes the features which were chosen by the emulator to be the most predictive when applied to the dark matter only simulation for the satellites model. It should be noted that the number of features per tree for the satellites model exceeds the total number of features, so is equivalent to using all features in a tree.}
\label{table:hyperparamchoicesnew}
\end{table*}
\endgroup

\subsection{Model Choice and Hyperparameter Optimisation}

As seen in works such as \cite{joMachineassistedSemisimulationModel2019}, \cite{lovellMachineLearningApproach2022}, \cite{kamdarMachineLearningCosmological2016b}, Extremely Randomized Trees provides a simple to implement yet very effective method for modelling the galaxy-halo connection, which we elect to use in this work. In order to improve the predictive ability of our pipeline, we train two separate models for centrals and satellites. This allows us to provide features to each model which are only relevant (e.g. host halo mass for satellites) and make use of the matched catalogue for centrals. As a result, we optimise the hyperparameters of these models separately.

Furthermore, we choose to treat the problem as a classification problem as in \cite{rodriguesHighfidelityReproductionCentral2023} and \cite{rodriguesExploringHalogalaxyConnection2025} to produce estimates of the probability distribution of the stellar mass as the output of our model, quantised in logarithmic space. This allows us to more accurately characterise the scatter in the stellar mass-halo mass relation, since regression would make an estimate of the mean of this distribution, which would narrow the overall scatter in stellar masses. 

The quantisation of the stellar masses used for the model gives rise to the number of stellar mass bins used as a hyperparameter to be optimised. This is in addition to the number of trees, the maximum depth that the trees are allowed to reach, the minimum number of samples at a node in the trees at which to split and the maximum number of input features to use in a tree. We also vary the amount of training data supplied to the model as increasing the training set improves the model but also increases the size of the model. These hyperparameters are detailed in Table \ref{table:hyperparamchoicesnew}. 

These hyperparameters give a parameter space from which we want to find our optimal combination of hyperparameters. We choose to emulate this parameter space using a Gaussian processes based emulator from \texttt{scikit-optimize} \citep{headScikitoptimizeScikitoptimize2022}. 

At a given point in the hyperparameter space, we train an Extremely Randomized Trees model using \texttt{scikit-learn} \citep{scikit-learn}. We then evaluate a metric $\mathcal{F}$ defined in Equation \ref{eq:metrmain} using the predictions from the models on the L1\_m8\_DMO validation set and the true values from the validation set in L1\_m8. For the computation of this metric, we sort the halos into 16 bins spaced equally in logarithmic space by their $M_{\rm{DM}}$ from $10^{11.5}\rm{M_{\odot}}$ to $10^{13}\rm{M_{\odot}}$. We select these limits to optimize the hyperparameters only in a region where there are a large number of galaxies that are well resolved. We choose to use $M_{\rm{DM}}$ as the dark matter property to bin the data due to its strong correlation with stellar mass while being relatively insensitive to baryonic feedback (compared with $V_{\rm{DM, peak}}$, for example). Each bin is then labelled from 1 to 16.

\begin{equation}
\label{eq:metrmain}
  \begin{aligned}
    \mathcal{F}=\;& \alpha\Delta(16)+\beta\Delta(50) + \gamma \Delta(84) \\
    +\;& \epsilon Q(16)+\zeta Q(50) + \eta Q(84)
  \end{aligned}
\end{equation}
\\
\begin{equation}
\label{eq:Delta}
    \Delta(N) = \log_{10} \left( \sqrt{\sum_{i=1}^{16}\frac{\delta_{\text{pred}}(N,i)^2}{\text{pcl}(M_{\star,\text{true}},N,i)^2}}\right)
\end{equation}
\\
\begin{equation}
\label{eq:littledeltaa}
    \delta_{\text{pred}}(N,i) = \text{pcl}(M_{\star,\text{pred}},N,i) - \text{pcl}(M_{\star,\text{true}},N,i)
\end{equation}
\\
\begin{equation}
\label{eq:Q}
    Q(N) = \log_{10} \left( \sqrt{\sum_{i=1}^{16}\frac{\delta_{\rm{quant}}(N,i)^2}{\text{pcl}(M_{\star,\text{true}},N,i)^2}}\right)
\end{equation}
\\
\begin{equation}
\label{eq:littledeltab}
   \delta_{\rm{quant}}(N,i) = \text{pcl}(M_{\star,\text{quant}},N,i) - \text{pcl}(M_{\star,\text{true}},N,i)
\end{equation}

In Equation \ref{eq:Delta}, we evaluate the errors in the percentiles for all $M_{\rm{DM}}$ bins. We define the $\text{pcl}(M_{\star}, N, i)$ function which computes the Nth percentile of $M_{\star}$ for galaxies in $M_{\rm{DM}}$ bin $i$. We define $\delta_{\text{pred}}(N,i)$ in Equation \ref{eq:littledeltaa}, which gives the difference in the Nth percentile stellar masses from the model predictions and true galaxies in  $M_{\text{DM}}$ bin $i$. For the predictions, we fit a cubic spline to the discrete probability distribution to give a continuous function from which to compute the percentiles.

We define $Q(N)$ in Equation \ref{eq:Q}, which gives an estimate of the error in our probability distributions due to their quantisation. This quantisation of the probability distribution estimates arises from two effects: the stellar masses are quantised as part of the model and the probability estimates for each stellar mass bin are quantised as a result of having a finite number of trees in the model. We estimate these quantisation errors by taking the true distribution of stellar masses within each $M_{\text{DM}}$ bin and binning it in stellar mass and in number density. This new quantised distribution then has different percentiles, from which we compute the errors in Equations \ref{eq:Q} \ref{eq:littledeltab} in a similar fashion to in Equations \ref{eq:Delta} and \ref{eq:littledeltaa}.

Equation \ref{eq:metrmain} linearly combines the six properties which are to be minimised with coefficients $\alpha$, $\beta$, $\gamma$, $\epsilon$, $\zeta$ and $\eta$. We decide to set $\alpha,\beta,\gamma=2$ and $\epsilon,\zeta,\eta=1$ in order to prioritise the minimisation of the errors in predictions over minimising the quantisation error.

We also penalise models that are too large. Since the trained models are available as data products, their usefulness to the wider community is minimal if they are prohibitively large. We monitor the memory used by the trained models during the hyperparameter optimisation and if the size exceeds 100GB, we set $\mathcal{F}$ to 100 (where the range of values that  $\mathcal{F}$ takes varies from approx -10 to 10 otherwise). 

At each point we employ a jackknife method by training five models each on 80\% of the data, removing a different 20\% of the data for each model. We then take the mean of the five values of $\mathcal{F}$ computed from each model's predictions. It should therefore be noted that the data fraction used to train each model is 80\% that given by the hyperparameter.

\begin{center}
    \begin{figure}[!h]
        \centering
    	\includegraphics[scale=1]{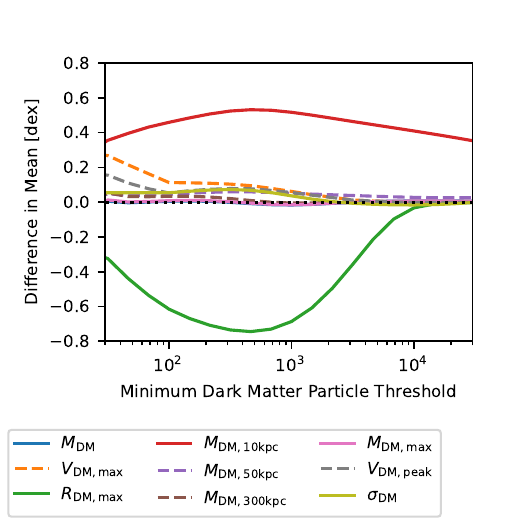}
    	\caption{The difference in the means of the distributions of the dark matter properties between L1\_m8 and L1\_m8\_DMO as a function of a minimum particle threshold applied, for all halos. This has been shown for a selection of features with high feature importance or substantial difference in distribution between the hydrodynamic and DMO simulations. Lines are shown variably as solid or dashed purely for ease of distinguishing. The zero line is shown in dotted black. All L1\_m8\_DMO properties have been adjusted to account for the difference in dark matter particle masses.}
        \label{fig:meandiffsdmo}
    \end{figure}
\end{center}

While previous works such as \cite{rodriguesExploringHalogalaxyConnection2025} have trained purely using hydrodynamic data, they note that a model trained like this would not necessarily produce reasonable results when applied to a DMO simulation, due to differences between them caused by baryonic feedback. In Figure \ref{fig:meandiffsdmo} we show how the means and standard deviations respectively differ between L1\_m8 and L1\_m8\_DMO as a function of the minimum dark matter particle threshold (which we have set to be 100) for a number of dark matter properties with either high correlation with stellar mass or substantially differing distributions between the hydrodynamic and DMO simulations. We see especially that $M_{\rm{DM, 10kpc}}$ and $R_{\rm{DM,max}}$
show substantial differences between the hydrodynamic and DMO runs even when making a stricter cut in dark matter particle number. The properties which have not been shown here have less than $0.2\;\rm{dex}$ difference between L1\_m8 and L1\_m8\_DMO at particle threshold of 100. $M_{\rm{DM, 10kpc}}$ especially never shows strong agreement between the hydrodynamic and DMO runs, which is likely due to the lack of AGN feedback in L1\_m8\_DMO, since this property is probing only the central-most part of the dark matter halo where AGN feedback is the most impactful. It would be possible to remove $M_{\rm{DM, 10kpc}}$ and make a stricter cut in dark matter particle number to reduce the differences between L1\_m8 and L1\_m8\_DMO. However this would reduce the stellar mass range over which the SMF is reasonable, as demonstrated in Figure \ref{fig:smfvspart}.

To optimise our pipeline's performance when applied to DMO simulations, we want to select hyperparameters which minimise the difference between the predictions from a DMO simulation and the halos in the hydrodynamic simulation. In light of this, we use the validation data set from L1\_m8\_DMO with which we optimise the models. This data set is 10\% of the halos in L1\_m8\_DMO with at least 100 dark matter particles. During the hyperparameter optimisation, we apply the model to the DMO validation set to produce a set of predictions. These predictions are used to compute $\rm{perc}(M_{\star,pred},N,i)$ when evaluating $\Delta(N)$ in the metric $\mathcal{F}$, effectively minimising the difference in what the model predicts from a DMO simulation and the ground truth hydrodynamic simulation. It should be noted that the validation set from L1\_m8 is still used to compute the $Q(N)$ terms as before, since the $Q(N)$ terms need to use the FLAMINGO stellar masses in their evaluation.

For the centrals model (hereby referred to as \texttt{DMO-opt-cent}), we make use of the matched set of halos from L1\_m8\_DMO. \texttt{SOAP} is able to match between hydrodynamic halos in L1\_m8 and DMO halos in L1\_m8\_DMO. This allows us to use the stellar mass from L1\_m8 as the training target parameter and the features of the halos in L1\_m8\_DMO as the input features for training. However, \texttt{SOAP} is only able to match central halos in L1\_m8 so we can only use this approach for \texttt{DMO-opt-cent} and not the model for predicting satellites.  We use 128 initial points for the emulator before it begins optimising, which we generate using a Sobol sequence implemented in \texttt{SciPy} \citep{2020SciPy-NMeth}. This gives more even coverage of our hyperparameter space when compared to pseudorandom generation of points. We then evaluate an additional 824 points using the emulator before selecting the optimal hyperparameter combination. 

For the satellites model (hereby referred to as \texttt{DMO-opt-sat}), we want to use only a set of input features which maximise the quality of the predictions made when applied to the dark-matter-only simulation. Therefore, we introduce a further 23 new binary hyperparameters, one for each dark matter halo feature to represent whether the model should be trained on the given dark matter feature. This allows the emulator to select which combination of input features produce the best predictions when applied to a DMO simulation run. Therefore, we expect this process to choose not to use features such as $R_{\rm{max}}$ and $M_{\rm{DM,10kpc}}$ that have a substantial difference in the DMO runs (see Figure \ref{fig:meandiffsdmo}). Along with the existing 6 parameters, we have a total of 29 parameters to optimise in our emulator. Given the increased size of our parameter space for \texttt{DMO-opt-sat}, we use 1024 random points before the Gaussian Processes emulator begins interpolating, again selected using a Sobol sequence to maintain even distribution across the parameter space. We then evaluate a further 1023 points using the emulator before selecting the optimal hyperparameters. Finally, we train a further 23 models with the selected hyperparameters and features except with one input feature used in the model if selected not to by the emulator and vice versa. This then gives a measure of whether including/excluding a given feature would improve the model. Finally, we train and validate a model with all of these feature adjustments included, which produced the best value of $\mathcal{F}$.

For \texttt{DMO-opt-sat}, we also need to apply a correction factor to a number of the dark matter properties in DMO simulations that it is applied to. L1\_m8\_DMO has an increased dark matter mass content compared with L1\_m8 in order to maintain the same total mass content. This manifests by an increased dark matter particle mass in L1\_m8\_DMO. Therefore, a property such as $M_{\rm{DM}}$ will be larger in the dark-matter-only run by the proportion of their dark matter particle sizes ($8.4 \times 10^8 / 7.06*10^8=1.19$). $M_{\rm{DM}}$, $M_{\rm{DM, 10kpc}}$, $M_{\rm{DM, 30kpc}}$, $M_{\rm{DM, 50kpc}}$, $M_{\rm{DM, 100kpc}}$, $M_{\rm{DM, 300kpc}}$, $M_{\rm{DM, 500kpc}}$, $M_{\rm{DM, 1000kpc}}$, $M_{\rm{DM, 3000kpc}}$, $M_{\rm{DM, max}}$, $M_{\rm{DM, host}}$ and $L_{\rm{DM}}$ in the dark-matter-only catalogues were thus scaled by $1/1.19=0.84$. Furthermore, the maximum of the circular velocity profile as it is computed by SOAP is given by Equation \ref{eq:vmax}. Therefore, $V_{\rm{DM, max}}$ and $V_{\rm{DM, peak}}$ must both be scaled by a factor of $\sqrt{0.84}=0.92$ in the dark-matter-only simulation. It should be noted that this correction factor does not resolve the lack of baryonic feedback in L1\_m8\_DMO. This correction is not necessary for \texttt{DMO-opt-cent} as that model is also trained on the dark matter content of L1\_m8\_DMO.

\begin{equation}
    \label{eq:vmax}
    v_{\rm{max}} = \sqrt{\frac{GM(\leq r)}{r}}
\end{equation}

 The new optimal hyperparameters are shown in Table \ref{table:hyperparamchoicesnew}. As expected, we find that the features with substantial differences between the hydrodynamic and DMO runs were not selected for \texttt{DMO-opt-sat}. Appendix \ref{ap:hyper} gives more detail on how the hyperparameters affect the metric $\mathcal{F}$ and on how the choice of input features affects $\mathcal{F}$. In the computation of $Q(50)$, we find much improved results when the number of trees in the model exceeds the number of stellar mass bins. We also find that $M_{\rm{DM,10kpc}}$ reduces the quality of the model's predictions the most and $\sigma_{\rm{DM}}$ improves the predictions the most. However, our final model does not use $\sigma_{\rm{DM}}$ as the improvement of $\mathcal{F}$ from the Sobol sequence does not indicate that it improves predictions; instead its high correlation causes models to use it in predictions instead of features such as $M_{\rm{DM,10kpc}}$ that would make predictions substantially worse. 

The pipeline with \texttt{DMO-opt-cent} used to predict for centrals and \texttt{DMO-opt-sat} used to predict satellites is hereby referred to as \texttt{DMO-opt}. This contrasts with a pair of models trained in using purely data from L1\_m8, collectively referred to as \texttt{HYDRO-opt}. The \texttt{HYDRO-opt} model does not use the matched halos from L1\_m8, instead training using dark matter features in the hydrodynamic simulation. Additionally, all dark matter features are used for training both the centrals and satellites models in \texttt{HYDRO-opt}. Further details of the \texttt{HYDRO-opt} model, including its performance on the test set from L1\_m8 are given in Appendix \ref{ap:hyper}.

\begingroup 
    \setlength{\tabcolsep}{10pt} 
    \renewcommand{\arraystretch}{1.5} 
    \setlength\extrarowheight{2pt}
\begin{table}[]
\centering
\begin{tabular}{|p{1.75cm}|p{1.5cm}|p{1.5cm}|}
\hline
Feature Name & Central Feature Importance & Satellite Feature Importance \\ \hline\hline
$V_{\rm{DM,peak}}$ & 0.151 & N/A \\ \hline
$M_{\rm{DM,50kpc}}$ & 0.144 & N/A \\ \hline
$M_{\rm{DM,100kpc}}$ & 0.104 & N/A \\ \hline
$M_{\rm{DM,30kpc}}$ & 0.103 & N/A \\ \hline
$z_{\rm{birth}}$ & 0.081 & N/A \\ \hline
$\sigma_{\rm{DM}}$ & 0.062 & N/A \\ \hline
$V_{\rm{DM,max}}$ & 0.062 & N/A \\ \hline
$M_{\rm{DM}}$ & 0.055 & 0.059 \\ \hline
$M_{\rm{DM,max}}$ & 0.047 & 0.155 \\ \hline
$M_{\rm{DM,300kpc}}$ & 0.041 & N/A \\ \hline
$M_{\rm{DM,500kpc}}$ & 0.035 & N/A \\ \hline
$M_{\rm{DM,1000kpc}}$ & 0.021 & 0.026 \\ \hline
$M_{\rm{DM,10kpc}}$ & 0.020 & N/A \\ \hline
$R_{\rm{DM,1/2}}$ & 0.016 & N/A \\ \hline
$z_{\rm{DM,peak}}$ & 0.010 & N/A \\ \hline
$M_{\rm{DM,3000kpc}}$ & 0.010 & 0.012 \\ \hline
$L_{\rm{DM}}$ & 0.008 & N/A \\ \hline
$R_{\rm{DM,max}}$ & 0.007 & N/A \\ \hline
$z_{\rm{DM, max}}$ & 0.006 & N/A \\ \hline
$c_{DM,200\rm{m,cent}}$ & 0.006 & N/A \\ \hline
$c_{DM,200\rm{c,cent}}$ & 0.006 & N/A \\ \hline
$V_{\rm{DM,COM}}$ & 0.006 & 0.010 \\ \hline
$V_{\rm{DM,ratio,sat}}$ & N/A & N/A \\ \hline
$M_{\rm{DM,host, sat}}$ & N/A & 0.011 \\ \hline
$R_{\rm{DM,ratio,sat}}$ & N/A & N/A  \\ \hline
\end{tabular}
\caption{The feature importances for \texttt{DMO-Opt}. The features are defined in Table \ref{table:features}. They have been ordered by the central feature importance. The satellite feature importances have been scaled down such that the mean feature importance is the same for centrals and for satellites to allow for more direct comparison.}
\label{table:importancesnew}
\end{table}
\endgroup

\begin{figure*}[!t]
    \centering
    \includegraphics[scale=1]{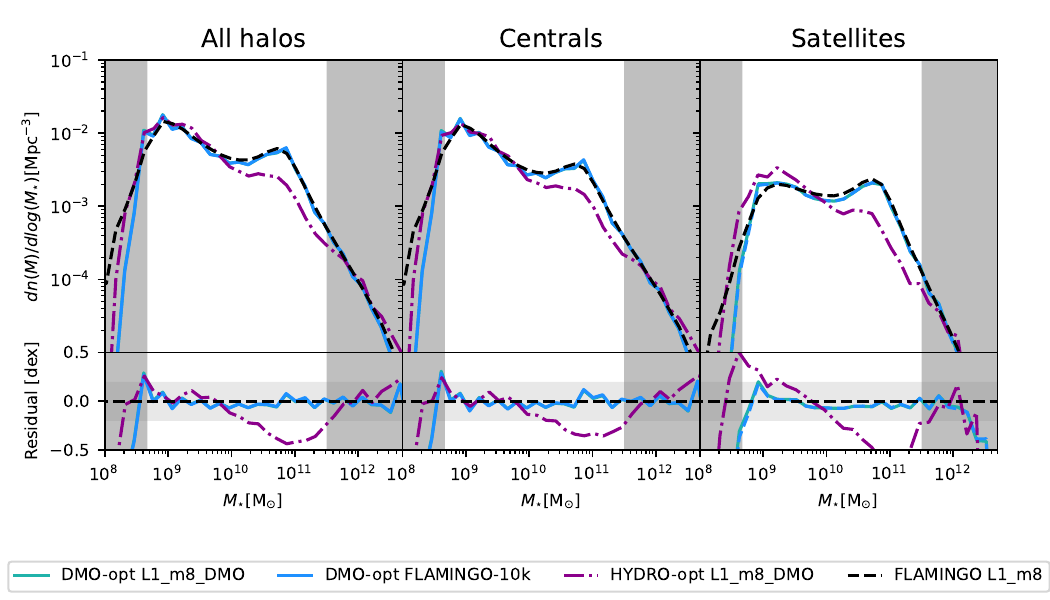}
    \caption{The stellar mass function shown for the predictions from \texttt{DMO-opt} (dashed cyan line) made using the test set from L1\_m8\_DMO. \texttt{DMO-opt} is a prediction pipeline optimised for its predictions when applied to a DMO simulation. We also show predictions made using the \texttt{HYDRO-opt} pipeline applied to that same test set. The \texttt{HYDRO-opt} pipeline has been trained only using data from the hydrodynamic L1\_m8 simulation. Finally we also show the true SMF from the L1\_m8 test set (black dotted line). The three columns correspond to the SMF for all halos, just centrals and just satellites respectively. The bottom row shows the residuals of the predictions compared with the SMF from the L1\_m8 test set. The light grey region in the residuals corresponds to the maximum difference between the FLAMINGO stellar mass function and the calibration data within the calibration range. The dark gray regions show the stellar mass ranges which are outside the calibration range. The predictions from the \texttt{DMO-opt} pipeline are within the 0.2\;dex range with which the L1\_m8 SMF differs from calibration data over the mass range where the SMF was calibrated and is reasonable. On the other hand, the predictions \texttt{HYDRO-opt} differ by around 0.5\;dex at $10^{11}M_{\odot}$.}
    \label{fig:smfhydrodmoopt}
\end{figure*}

\begin{figure*}[!t]
    \centering
    \includegraphics[scale=1]{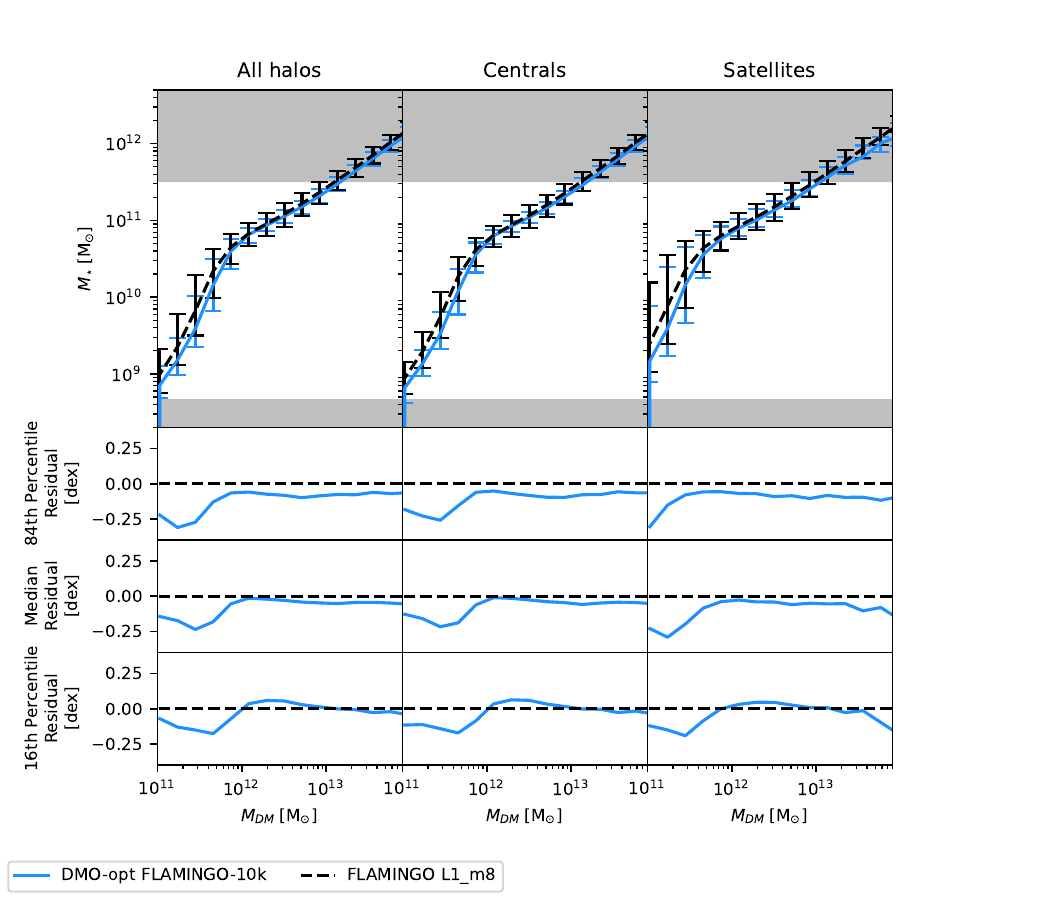}
    \caption{The percentiles of the stellar masses as a function of the dark matter mass, shown for the predictions from \texttt{DMO-opt} on FLAMINGO-10k (solid blue line). Also shown are the true percentiles from L1\_m8 (dashed black line). The columns and grey regions are the same as in Figure \ref{fig:smfhydrodmoopt}. In the top panel, the lines correspond to the medians and the error bars correspond to the 16th and 84th percentiles. The rows below then depict the residuals of the percentiles with respect to L1\_m8. For $M_{\rm{DM}}>10^{12}\rm{M_{\odot}}$ halos, we see strong agreement to within $\approx 0.1\;\rm{dex}$ for all percentiles. This increases to $\approx 0.25\;\rm{dex}$ for lower mass halos.}
    \label{fig:smhmsecond}
\end{figure*}

\begin{figure*}[!]
    \centering
    \includegraphics[scale=1]{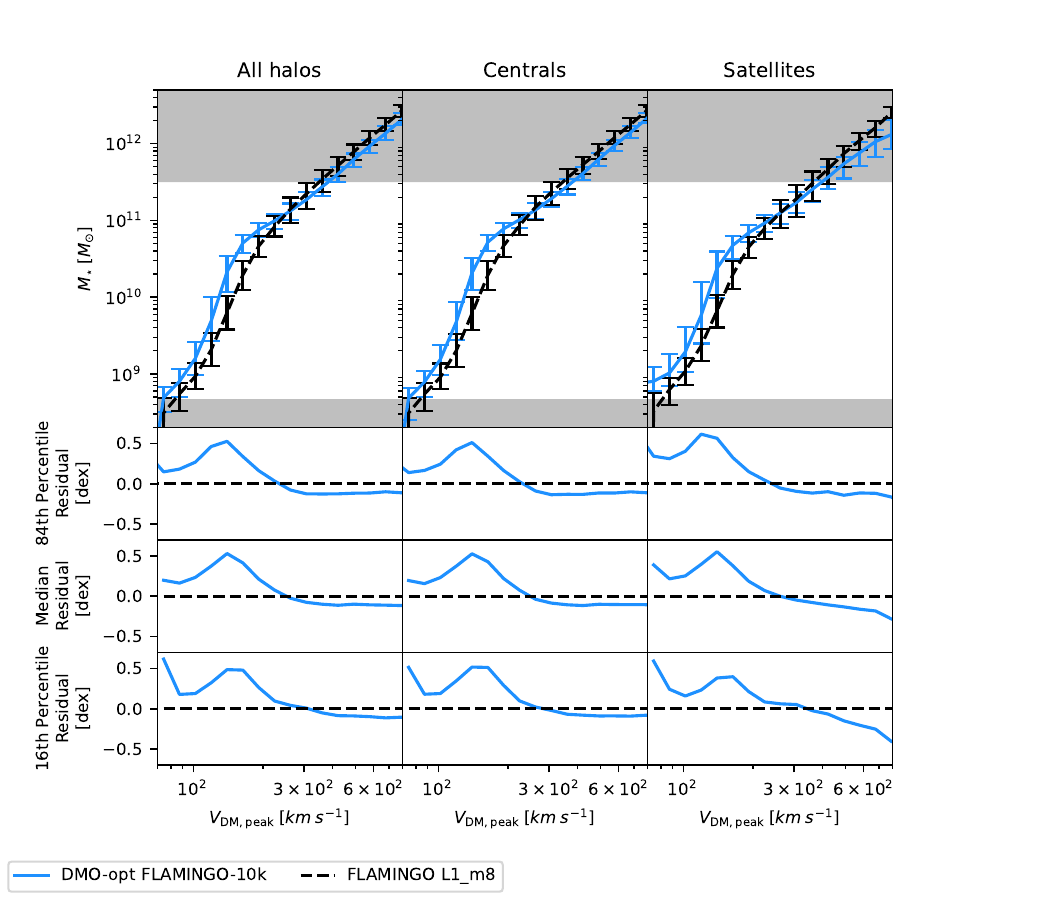}
    \caption{Similar to Figure \ref{fig:smhmsecond}, the percentiles of stellar mass as a function of $V_{\rm{DM,peak}}$ for predictions from \texttt{DMO-opt} on FLAMINGO-10k (blue solid line). Also shown are the percentiles of the true stellar masses in L1\_m8 (black dashed line). In the top panel, the lines correspond to the medians and the error bars correspond to the 16th and 84th percentiles. The rows below then depict the residuals of the percentiles with respect to L1\_m8. The values of $V_{\rm{DM,peak}}$ for the DMO simulation predictions have been rescaled to account for the higher dark matter content in the DMO simulations. We see reasonable agreement for $V_{\rm{DM,peak}}>300\rm\;km\;s^{-1}$ halos with mostly less than 0.2\;dex difference, however for $V_{\rm{DM,peak}}\approx 150\rm\;km\;s^{-1}$ we see all percentiles overestimated by around 0.5\;dex.}
    \label{fig:smvpeaksecond}
\end{figure*}

\begin{figure*}[!t]
    \centering
    \includegraphics[scale=1]{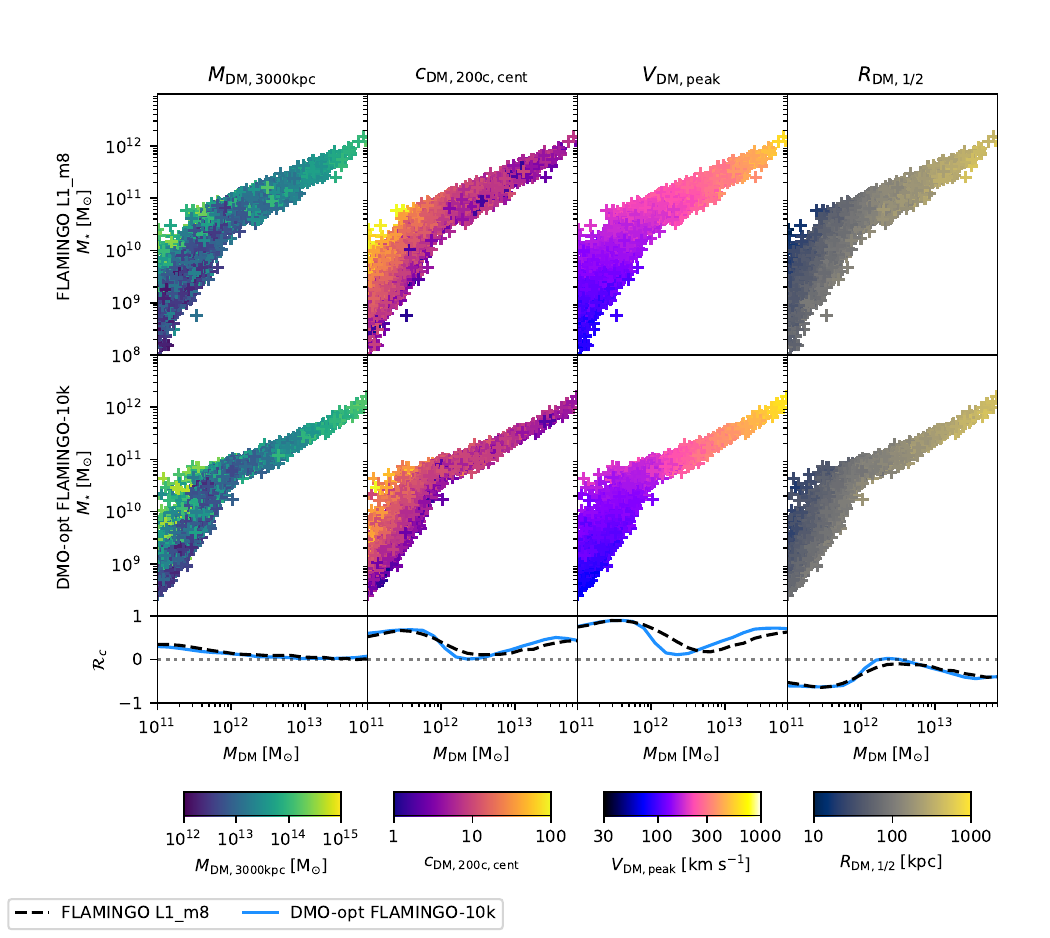}
    \caption{The stellar mass-dark matter mass relation coloured by four different other dark matter properties in each column: $M_{\rm{DM,3000kpc}}, c_{\rm{DM,200c,cent}},V_{\rm{DM,peak}}, R_{\rm{DM,1/2}}$. The top row shows these relations for the galaxies in L1\_m8. The middle row shows these relations for the predictions made from the \texttt{DMO-opt} model from the FLAMINGO-10k DMO simulation. The top two rows are shown for a random sample of 10,000 central galaxies from each simulation. To produce the galaxies for FLAMINGO-10k, the distributions predicted by \texttt{DMO-opt} were directly sampled from. The bottom row shows the Spearman's rank correlation between stellar mass and the dark matter property used to colour the distributions as a function of dark matter mass e.g. the bottom-left panel shows the Spearman's rank correlation coefficient between $M_{\star}$ and $M_{\rm{DM,3000kpc}}$ as a function of $M_{\rm{DM}}$. This is shown both for the L1\_m8 galaxies (dashed black line) and the FLAMINGO-10k predictions using \texttt{DMO-opt} (solid blue line). We see strong agreement in the correlation over all mass ranges for $M_{\rm{DM,3000kpc}}$, $c_{\rm{DM,200c,cent}}$, and $R_{\rm{DM,1/2}}$. There is greater discrepancy in the correlation for $V_{\rm{DM,peak}}$, due to $V_{\rm{DM,peak}}$ having more substantial differences between the hydrodynamic and DMO simulations.}
    \label{fig:trirel}
\end{figure*}

\begin{figure*}[!t]
    \centering
    \includegraphics[scale=1]{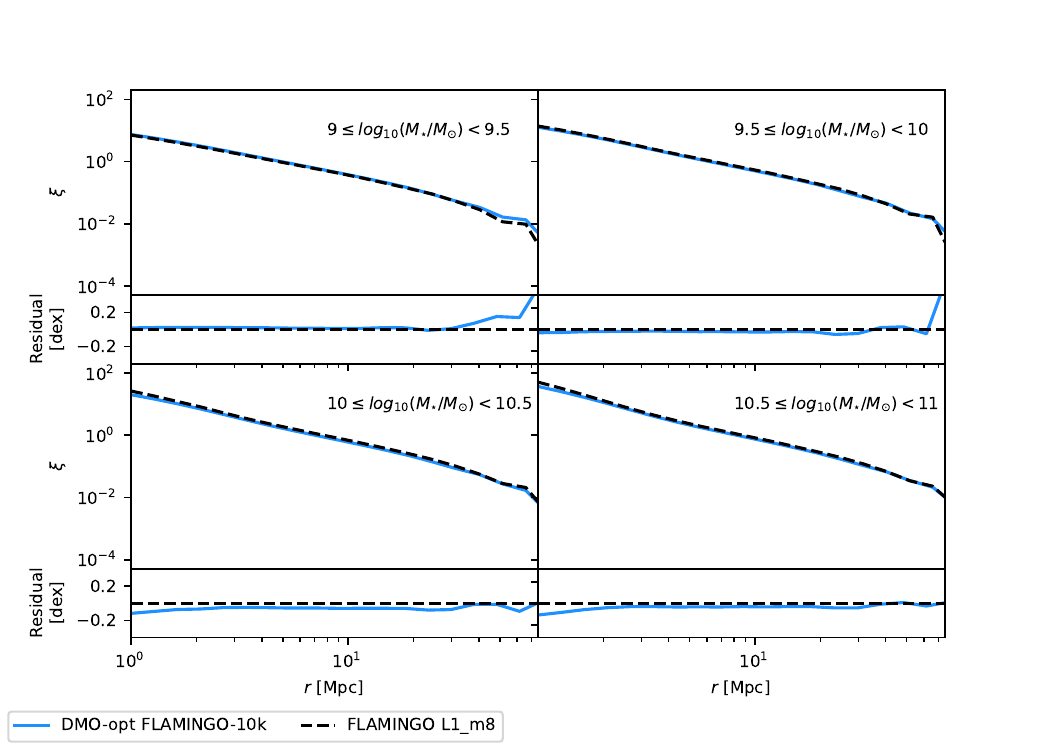}
    \caption{The 2-point correlation functions for the predictions from \texttt{DMO-opt} from FLAMINGO-10k (solid blue line) for four stellar mass bins. In order to produce this, the distributions predicted by \texttt{DMO-opt} were directly sampled from. Also shown is the correlation function for L1\_m8 (dashed black line). Both correlations are computed by taking the most bound particle of the halo to be the position of the halo. We see strong agreement of less than 0.1\;dex over almost all length scales for all four stellar mass bins.}
    \label{fig:corp2}
\end{figure*}

\section{Results}
\label{sec:results}

\subsection{Results From DMO-opt applied to Test Set}

We train \texttt{DMO-opt} on our training set,  using 80\% of the data fraction selected by the optimiser to account for the jackknife method used in the hyperparameter optimisation. Table \ref{table:importancesnew} lists the feature importances for all of the input features in \texttt{DMO-opt}, ordered by the feature importance for the centrals model. The feature importances for the satellites model have been rescaled so that the mean feature importance for the two models is the same. For a given feature, the importance is computed as the mean over all trees of the accumulative reduction in the Gini impurity produced by inclusion of the feature. This is then normalised so that the sum of all feature importances is one. The Gini impurity $H$ at a given node $Q_m$ in a tree is defined in Equation \ref{eq:gini}, where $p_{mk}$ is the proportion of observations of class $k$ at node $Q_m$\footnote{\href{https://scikit-learn.org/stable/modules/tree.html\#mathematical-formulation}{https://scikit-learn.org/stable/modules/tree.html\#mathematical-formulation}}.

\begin{equation}
    \label{eq:gini}
    H(Q_m) = \sum_k p_{mk}(1-p_{mk})
\end{equation}

We can see that $V_{\text{DM,peak}}$ is the most important feature for centrals, which is unsurprising given its strong correlation with stellar mass and its popularity as a choice for the dark matter feature in SHAM. We also see high importance in centrals for $M_{\rm{DM, 30kpc}}$, $M_{\rm{DM, 50kpc}}$ and $M_{\rm{DM, 100kpc}}$, with reduced importance for the larger spheres. This is also unsurprising as smaller spheres will moreso characterise the region of the halo which is influential on the galaxy formation \citep[e.g.][]{zavala2016}. It is for the same reason that measures of the maximum of the circular velocity are popular choices for SHAM as they are typically reached at around one-tenth of the halo radius \citep{chaves-monteroSubhaloAbundanceMatching2016}. For satellites, we see that $M_{\rm{DM,max}}$ is substantially more important than it is for centrals and is the most important feature for satellites. It should be noted that the satellites feature importances are rescaled to give the same mean feature importance so that the importances for centrals and satellites can be directly compared.

We then apply the \texttt{DMO-opt} to L1\_m8\_DMO and to FLAMINGO-10k, which has the same resolution as was used for training but a substantially larger box size. We also show the \texttt{HYDRO-opt} pipeline which was optimised and trained using all features and only using hydrodynamic data. Further details of the optimisation and performance when applied to hydrodynamic data of the \texttt{HYDRO-opt} pipeline are given in Appendix \ref{ap:smhm}. Also shown is the SMF from L1\_m8 and the dark grey region corresponds to regions outside of which the L1\_m8 SMF was not calibrated to observational data and thus should not be trusted. All SMFs are shown only for halos with 100+ particles. For the predictions made from \texttt{DMO-opt} from both DMO simulations, we see strong agreement between the predictions and the true SMF over the entire stellar mass range in which the SMF was calibrated and is reasonable, with residuals less than 0.1\;dex. For comparison's sake we show a light-grey region in the residuals which corresponds to 0.2\;dex, the approximate maximum deviation of the FLAMINGO SMF from the calibration data within the calibration region. We therefore permit 0.2\;dex as the maximum acceptable deviation of the predictions of our models from L1\_m8. On the other hand, the \texttt{HYDRO-opt} SMF from L1\_m8\_DMO agrees very poorly with the L1\_m8 SMF, especially around $10^{11}\;\rm{M_{\odot}}$. The residual is approximately 0.4 dex in this region, greater than this threshold. For satellites this is even more pronounced, with deviations of over 0.5\;dex. Given that $M_{\rm{DM,10kpc}}$ correlates more strongly with $M_{\star}$ for satellites than for centrals (as seen in Figure \ref{fig:corrplot}) and $M_{\rm{DM,10kpc}}$ shows the one of the greatest disparities between L1\_m8 and L1\_m8\_DMO (as seen in Figure \ref{fig:meandiffsdmo}), this is not surprising. This demonstrates that a model trained purely with data from hydrodynamic simulations with no optimisation for its performance on DMO simulations will not perform well in that case.

In Figure \ref{fig:smhmsecond}, we show the stellar mass percentiles in the predictions from \texttt{DMO-opt} on FLAMINGO-10k as a function of $M_{\rm{DM}}$. The percentiles for the predictions from \texttt{DMO-opt} were computed from the total discrete distribution by interpolating between the bins. We also show these for the galaxies in L1\_m8. We see strong agreement in the predictions using FLAMINGO-10k, with all percentiles showing residuals less than $0.1\;\rm{dex}$ above $M_{\rm{DM}}=10^{12}\rm{M_{\odot}}$. However, there is increased disparity in all percentiles for both centrals and satellites below this mass range, with differences up to $0.3 \;\rm{dex}$. 

In Figure \ref{fig:smvpeaksecond}, we do not see such good agreement in the percentiles of stellar mass as a function of $V_{\rm{DM,peak}}$. Above $V_{\rm{DM,peak}}=200\rm{km\;s^{-1}}$, we see residuals in the percentiles of stellar mass in the predictions from FLAMINGO-10k mostly below $0.2 \;\rm{dex}$. However, the residuals are much greater below this point. This is unsurprising as there is substantial difference in the distributions of stellar mass as a function of $V_{\rm{peak}}$ between L1\_m8 and L1\_m8\_DMO, especially for lower mass halos.

In Figure \ref{fig:trirel}, we show the dark matter mass-stellar mass relation in four columns coloured by $M_{\rm{DM,3000kpc}}$, $c_{\rm{DM,200c,cent}}$, $V_{\rm{DM,peak}}$, $R_{\rm{DM,1/2}}$ respectively. The top row shows these relations for the galaxies in L1\_m8 and the middle row shows this for the predictions made from \texttt{DMO-opt} from FLAMINGO-10k. The bottom row shows the Spearman's rank correlation coefficient between stellar mass and the dark matter feature used to colour as a function of the dark matter mass. We see good agreement in the correlations over all dark matter mass ranges for $M_{\rm{DM,3000kpc}}, c_{\rm{DM,200c,cent}}, R_{\rm{DM,1/2}}$. The agreement is less good for $V_{\rm{DM,peak}}$ at higher masses, although $V_{\rm{DM,peak}}$ is a feature with substantial disparity between the hydrodynamic and DMO simulations. It should be noted that these distributions could show strong agreement in correlation while not being similar distributions (e.g. by having a constant offset in stellar mass), so should be considered in combination with the \texttt{DMO-opt} predictions of the median stellar mass from FLAMINGO-10k compared with the median stellar mass in L1\_m8 shown in Figure \ref{fig:smhmsecond}. The relation shown in the second column has been demonstrated to exist in other simulations such as in COLIBRE in \cite{wangGravitationalPotentialDrives2026}. The ability of \texttt{DMO-opt} to reproduce the correlation between concentration and stellar mass demonstrates our models' ability to model the multidimensional factors that influence galaxy formation.

In Figure \ref{fig:corp2} we show the correlation functions using the predictions made from \texttt{DMO-opt} on FLAMINGO-10k and the true correlation functions from L1\_m8. These are shown for 4 stellar mass bins, so to produce the FLAMINGO-10k correlations, the probability distributions from \texttt{DMO-opt} were directly sampled from. The correlations were computed using \texttt{treecorr} \citep{jarvisSkewnessApertureMass2004} in 20 bins of separation from 1\;Mpc to 100\;Mpc. They were computed using $20\times$ the number of galaxies in the lowest stellar mass bin (which has the largest number of the four bins) as the number of randoms. We see strong agreement between our predictions from FLAMINGO-10k and L1\_m8 over almost all length scales for all bins with residuals less than 0.1\;dex. We do see slight disparity at large length scales for the two lowest stellar mass bins. In the largest two stellar mass bins, we see a slight increased disparity at the smallest length scales, although this barely exceeds 0.1\;dex. This is a significant result given the use case of these catalogues for galaxy clustering; works such as \cite{dvornikKiDS1000CombinedHalomodel2023} make use of similar correlation functions in stellar mass bins \citep[although the correlation functions computed in ][ are projected 2D correlations rather than the 3D correlations computed here]{dvornikKiDS1000CombinedHalomodel2023}. This demonstrates that our model is able to accurately reproduce measurements of large scale structure despite only making predictions for individual halos.

\section{Conclusions}
\label{sec:conclusions}

In this work we explore the use of machine learning as a method for modelling the stellar mass probability distribution of individual FLAMINGO halos using their dark matter properties as inputs. Machine learning provides a method of modelling the complex relation between many properties of dark matter halos computed only using the dark matter particles that comprise them and the stellar mass of galaxies that reside within them while making minimal assumptions in the process. This is in contrast to Subhalo Abundance Matching which typically only uses a single or small number of dark matter features, and Semi-Analytic Models which require a greater number of theoretical assumptions in their galaxy formation models. 

By binning in stellar mass, we can treat each bin as a class for an Extremely Randomized Trees classifier. The combination of the predictions of the trees in the classifier then gives a prediction for the stellar mass probability distribution function, with the probability of each bin being the fraction of trees that predict that bin.

We train two classifiers, one to predict for centrals and one to predict for satellites. The hyperparameters of these models are optimised using a Gaussian Processes based emulator to reproduce the median, 16th and 84th percentile stellar mass in 16 dark matter mass bins. The metric used for hyperparameter optimisation also includes estimation of the error caused by the quantisation from the stellar mass binning and from having a finite number of trees in the models.

The centrals model makes use of a matched catalogue between L1\_m8 and L1\_m8\_DMO, the highest resolution hydrodynamic simulation in FLAMINGO and the equivalent DMO simulation with the same initial conditions. Due to resolution constraints it is only possible to match central halos in FLAMINGO, so cannot be used for satellites. Therefore, for the satellites model, we use the hyperparameter optimisation emulator to select which dark matter features from the hydrodynamic simulation should be used and which should not. The models are then validated against a set of halos taken from L1\_m8\_DMO and the percentiles of the stellar mass as a function of dark matter mass were then compared with that from a validation set from L1\_m8.

We apply these models to halo catalogues from L1\_m8\_DMO and FLAMINGO-10k, a DMO simulation with the same resolution as L1\_m8\_DMO but with a larger 2.8\;Gpc box side length. This pipeline (entitled \texttt{DMO-opt}) is shown in Figure \ref{fig:smfhydrodmoopt} to be effective at reproducing the stellar mass function when applied to an unseen set of halos. We find the predicted SMF to differ from the L1\_m8 SMF by substantially less than the 0.2\;dex deviation of the L1\_m8 SMF from the calibration data. For comparison, we also train a pair of models (entitled \texttt{HYDRO-opt}) which uses only data from the hydrodynamic simulation L1\_m8 and uses all dark matter features for training. \texttt{HYDRO-opt} performs much more poorly when applied to DMO halos due to differences between the hydrodynamic and DMO simulations as a result of baryonic feedback.

In Figure \ref{fig:smhmsecond}, we demonstrate that our models perform well at reproducing the stellar mass-dark matter mass relationship over a large range of dark matter mass, although there is increased disparity up to 0.25\;dex in the 16th, 50th and 84th percentile stellar mass below $M_{\rm{DM}}$ of $10^{12}\;\rm M_{\odot}$, corresponding to roughly 1000 dark matter particles. By reproducing the 16th and 84th percentiles accurately, we demonstrate that our models are able to reproduce the scatter in the stellar mass-dark matter mass relationship over a large range of masses. The models perform less well when applied to the stellar mass-$V_{\rm{DM,peak}}$ relation, as shown in Figure \ref{fig:smvpeaksecond}. This is unsurprising as $V_{\rm{DM,peak}}$ shows substantial disparity between the hydrodynamic and DMO simulations. 

In Figure \ref{fig:trirel}, we study the correlations between deviations from the stellar mass-dark matter mass relation and four other dark matter features, $M_{\rm{DM,3000kpc}}$, $c_{\rm{DM,200c,cent}}$, $V_{\rm{DM,peak}}$ and $R_{\rm{DM,1/2}}$. We found our model was able to reproduce these correlations effectively for all features except for $V_{\rm{DM,peak}}$, for the same reason as previously mentioned. This demonstrates our model's ability to model the connection between several dark matter properties and their influence on galaxy formation.

Finally, in Figure \ref{fig:corp2} we demonstrate that the catalogue produced for FLAMINGO-10k reproduces the 2-point correlation functions in four stellar mass bins, which is a similar quantity as is used in works estimating cosmological parameters through galaxy clustering \citep[e.g ][]{dvornikKiDS1000CombinedHalomodel2023}. This demonstrates that our model is able to reproduce measures of large scale structure despite only making predictions for individual halos.

Many other similar works did not use the binning and classification method which we employed. By producing probability distributions of the stellar mass, our models are able to more accurately capture the scatter in the stellar mass-halo mass relation, whereas works such as \cite{hausenRevealingGalaxyHalo2023} note their model's inability to capture this scatter. Some works such as \cite{rodriguesHighfidelityReproductionCentral2023} used a similar binning process as in this work. However, we included the number of bins as a hyperparameter which we optimised for, whereas that work chose to use 50 bins for all the baryonic properties estimated by their model. However, this scatter does contain an irreducible scatter due to the inherently stochastic nature of the FLAMINGO simulations \citep{borrowIrreducibleScatter}. Characterising this scatter as part of our models is beyond the scope of this work.

Using FLAMINGO does give our models an advantage over all other works using other simulations as FLAMINGO has a particularly large number of particles in a simulation run to $z=0$. This gives us a set of training halos orders of magnitude greater than those used to train similar models. 

Another advantage of our pipeline gained from our hyperparameter optimisation was the selection of features to train on for the satellites model. Works such as \cite{lovellMachineLearningApproach2022} were able to purely mitigate the differences in hydrodynamic and DMO runs by training on the dark matter properties of matched halos in a DMO run, however this then requires that the training set is restricted to halos that can be matched. For the FLAMINGO simulations, this would have required restricting our training set to just central halos, whereas through our methodology we are able to train and make predictions for satellite subhalos as well.

Our models are only able to make predictions for the probability distributions of stellar masses. Other works such as \cite{rodriguesExploringHalogalaxyConnection2025} produced models which were able to predict other properties such as star formation rates. A natural extension of this work would therefore be to train additional models to predict the probability distributions of other baryonic properties in FLAMINGO. Another natural extension of this work would be to train a model using an intermediate resolution FLAMINGO simulation as there are also simulations with variations to cosmological parameters at this resolution, thus allowing for tests of sensitivity of the predictions to the cosmological parameters.

We present the catalogues of stellar mass estimates from the FLAMINGO-10k simulation as a data product of this work, which will be made available as part of the FLAMINGO data release on publication. Furthermore, we make the models trained in this work available on request so that they can be applied to other simulations not included as part of this work.

\section*{Acknowledgments}

This work was supported by the Science \& Technology Facilities Council [Grant Reference No ST/W006790/1]. This work used the DiRAC@Durham facility managed by the Institute for Computational Cosmology on behalf of the STFC DiRAC HPC Facility (www.dirac.ac.uk). The equipment was funded by BEIS capital funding via STFC capital grants ST/P002293/1, ST/R002371/1 and ST/S002502/1, Durham University and STFC operations grant ST/R000832/1. DiRAC is part of the National e-Infrastructure. The authors acknowledge the Comet HPC facility at Newcastle University for providing computational resources that have contributed to the research results reported within this paper. This publication is based upon work from COST Action CA21136 – “Addressing observational tensions in cosmology with systematics and fundamental physics (CosmoVerse)”, supported by COST (European Cooperation in Science and Technology). MA is supported by the UK Science and Technology Facilities Council (STFC) under grant number ST/Y002652/1 and the Royal Society under grant number ICAR1231094. MMR is supported by the Royal Society International Exchanges Global Programme, reference number IES{\textbackslash}R1{\textbackslash}241442.

\bibliographystyle{apsrev4-1}

\bibliography{oja_template}

\newpage
\begin{appendix}

\begin{center}
    \begin{figure*}[]
        \centering
    	\includegraphics[scale=1]{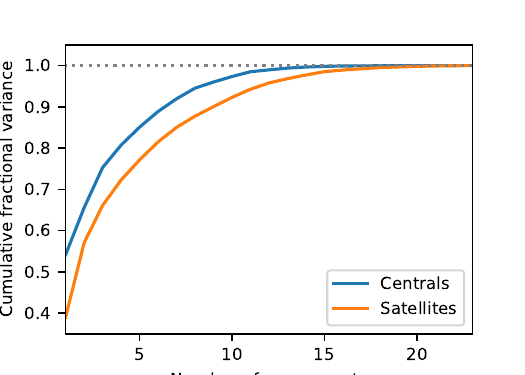}
    	\caption{The fractional cumulative variance of the principal components as a function of the number of components used, using the training data in L1\_m8.  The line of cumulative fractional variance of 1 is shown in dotted grey. The cumulative fractional variance gradually approaches 1 as the number of principal components is increased. This indicates that a similar number of principal components to the number of features would be needed for minimal information loss through PCA.}
        \label{fig:pca}
    \end{figure*}

    \begingroup 
        \setlength{\tabcolsep}{10pt} 
        \renewcommand{\arraystretch}{1.5} 
        \setlength\extrarowheight{2pt}
    \begin{table*}[]
    \centering
    \begin{tabular}{|l|l|l|l|}
    \hline
    Hyperparameter             & Full Range Explored & Central Optimal Value & Satellite Optimal Value \\ \hline\hline
    Number of Trees & 10-5000 & 1457 & 4189 \\ \hline
    Number of Features per Tree & 1-22 (Central) 1-23 (Satellite) & 20 & 23 \\ \hline
    Maximum Depth of Trees & 5-100 & 39 & 75 \\ \hline
    Minimum Samples to Split &2-1000 & 719 & 738 \\ \hline
    Number of Stellar Mass Bins &10-1000 & 723 & 322 \\ \hline
    Data Fraction & $10^{-5}$-$10^{0}$ (sampled logarithmically) &  0.334 & 0.455 \\ \hline
    \end{tabular}
    \caption{The ranges of hyperparameters explored and the optimal values found for \texttt{HYDRO-opt}. The final two columns show the optimal hyperparameter values for the models for centrals and satellites respectively.}
    \label{table:hyperparamchoices}
    \end{table*}
    \endgroup

    \begin{figure*}[!t]
        \centering
    	\includegraphics[scale=1]{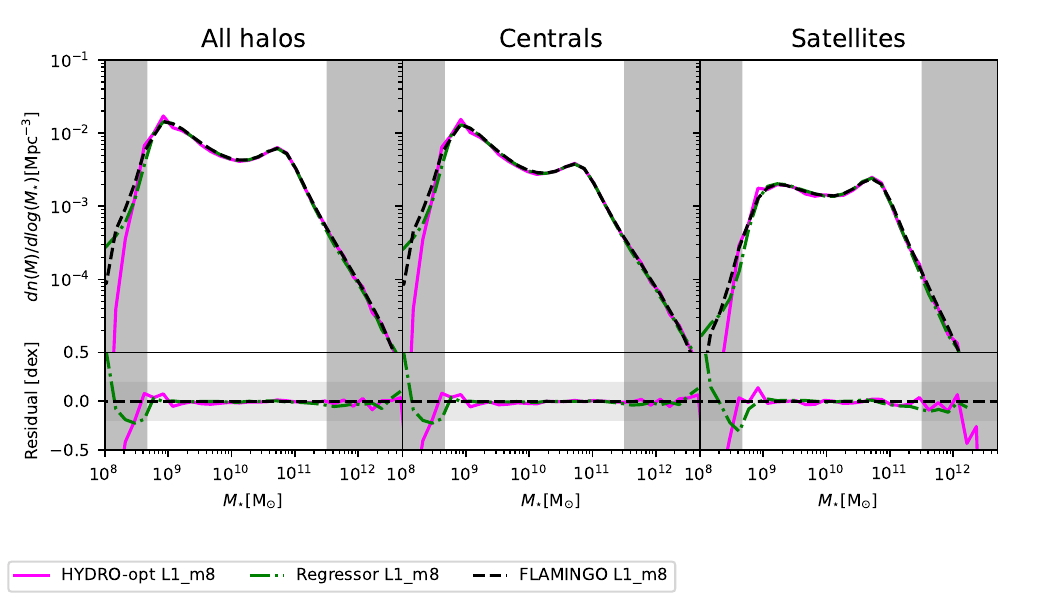}
    	\caption{The stellar mass function of the FLAMINGO galaxies in the L1\_m8 test data set (black dashed line), for the \texttt{HYDRO-opt} pipeline applied to that test data set (magenta solid line) and for the regressor applied to the test data set (green dash-dot line). The columns are for all galaxies, just central halos and just satellite subhalos respectively. The light grey region in the residual plots shows the maximum difference within the calibration stellar mass range by which the L1\_m8 SMF differs from the calibration data. The dark grey regions show regions outside of the calibration range. The bottom row shows the residuals of the predictions from the two pipelines compared with the SMF from the true FLAMINGO galaxies.\newline\newline}
        \label{fig:smf-p1}
    \end{figure*}

        \begin{figure}[!t]
        \centering
    	\includegraphics[scale=1]{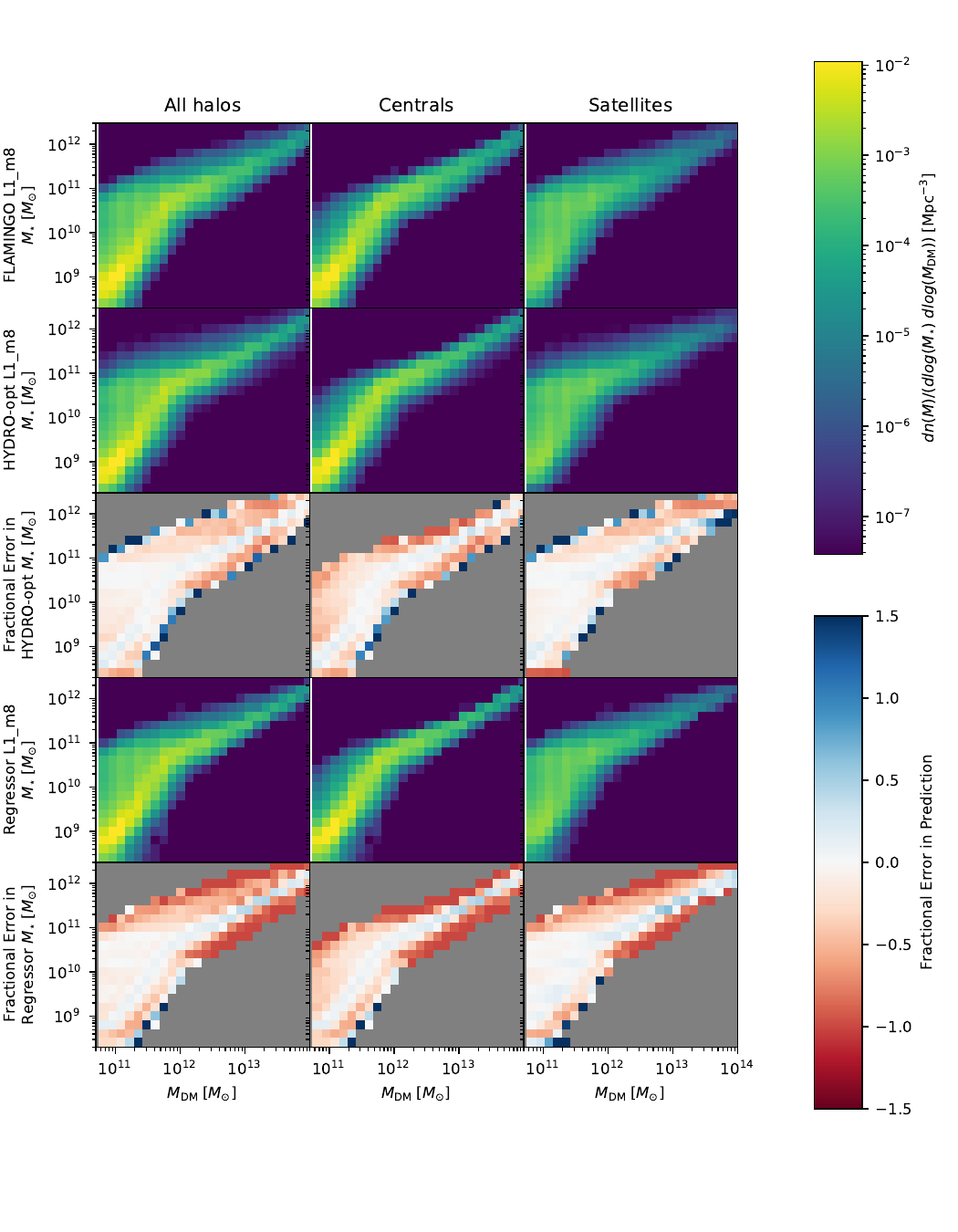}
    	\caption{2D plots of the number density of galaxies with a given stellar mass and dark matter mass. The three columns show this for all halos, just centrals and just satellites respectively. The first row shows the number density of galaxies for the FLAMINGO L1\_m8 simulation. The second row shows the number density from the predictions made with the \texttt{HYDRO-opt} pipeline applied to L1\_m8. The third row shows the fractional error in these predictions when compared with the FLAMINGO L1\_m8 galaxies. Regions in grey correspond to regions where there are no galaxies in the FLAMINGO L1\_m8 simulation. The fourth row shows the number density from the predictions made with the regressor models applied to L1\_m8. The final row shows the fractional error in the predictions made from the regressors when compared with the FLAMINGO L1\_m8 galaxies.}
        \label{fig:smhmfull}
    \end{figure}

    \begin{figure}[!t]
        \centering
    	\includegraphics[scale=1]{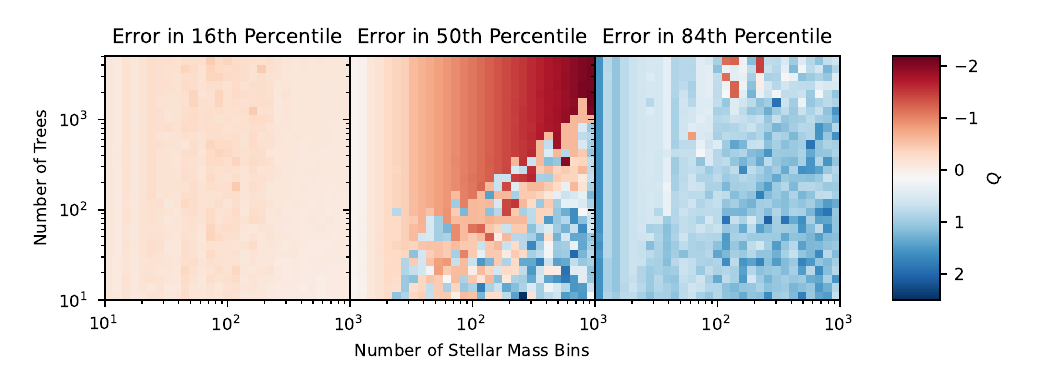}
    	\caption{Quantisation errors due to the number of stellar mass bins and number of trees. The panels from left to right show $Q(16)$, $Q(50)$ and $Q(84)$ respectively, defined in Equation \ref{eq:Q}. The colour scale is such that regions in red correspond to lower quantisation error and regions in blue to higher error. We find in $Q(50)$ that having more trees than stellar mass bins substantially reduces this quantisation error.}
        \label{fig:quanterror}
    \end{figure}
    \begin{figure}[!t]
        \centering
    	\includegraphics[scale=1]{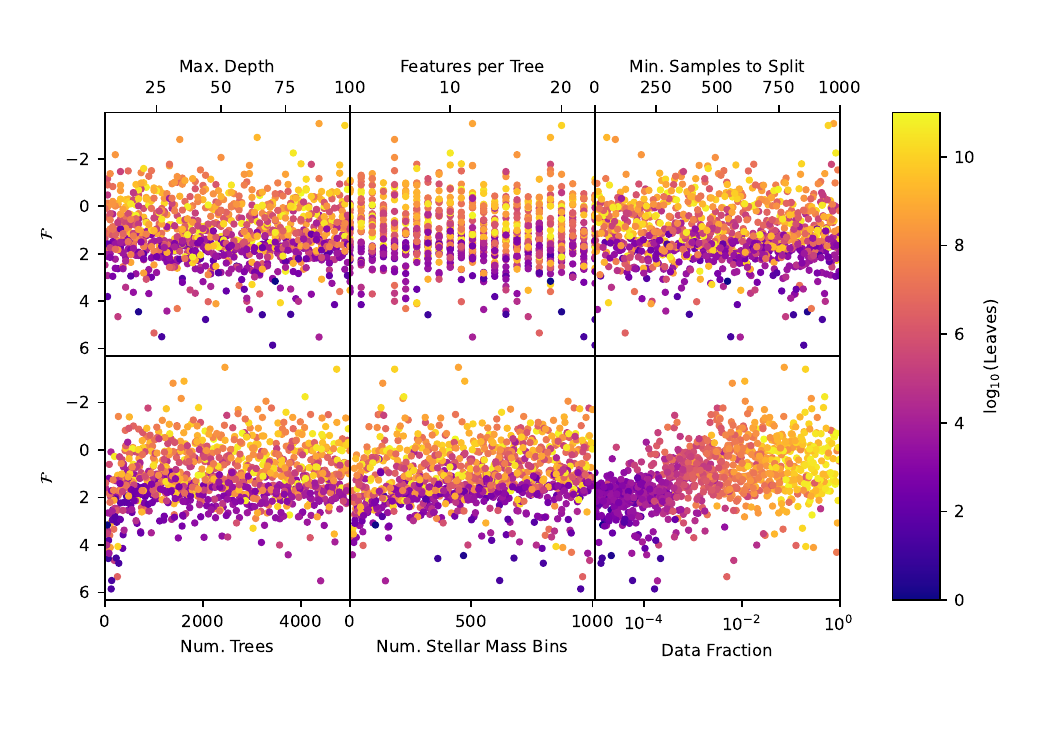}
    	\caption{Scatter plots of the 1024 points generated with a Sobol sequence used to initialise the emulator used for the hyperparameter optimisation of the \texttt{DMO-opt-sat} model. The top three panels from left to right show the loss function $\mathcal{F}$ against the maximum depth of the trees, the maximum number of features to use per tree and the minimum number of samples to split on. The bottom three panels from left to right show $\mathcal{F}$ against the number of trees in the model, the number of stellar mass bins and the fraction of the training data set used to train the model. The points are coloured by the average size of the models produced, measured by the total number of leaves in the trees. Here we exclude any points for which $\mathcal{F}=100$ due to the model exceeding our maximum memory cap.}
        \label{fig:sobolscatter}
    \end{figure}

    \begin{figure}[!t]
        \centering
    	\includegraphics[scale=1]{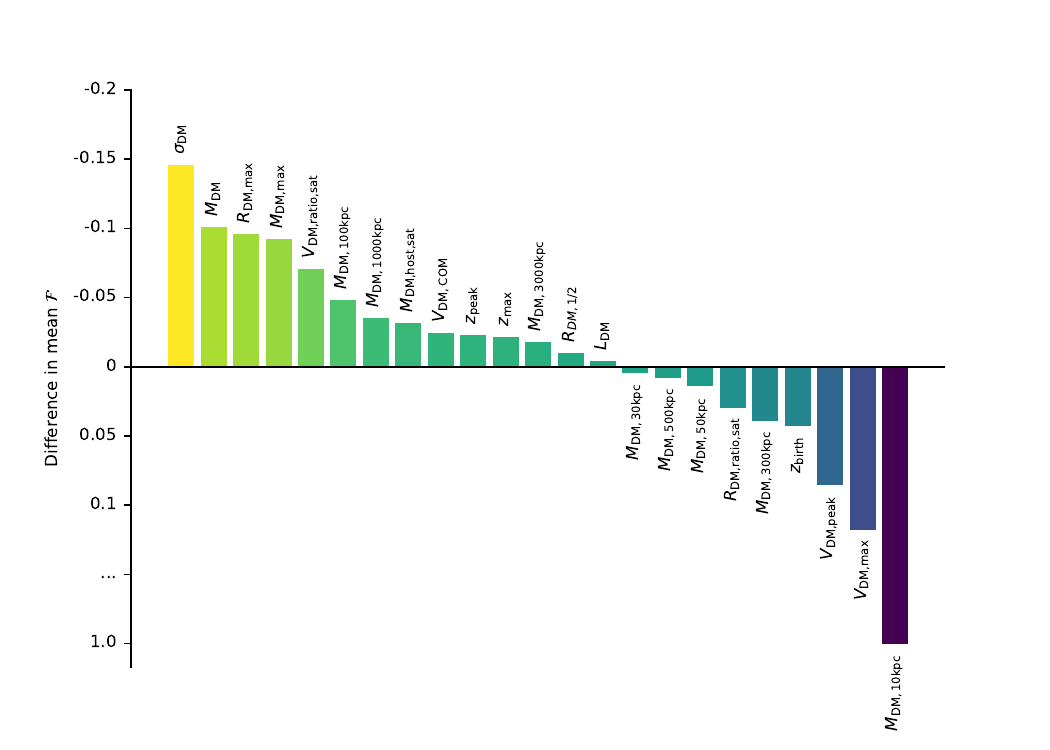}
    	\caption{The mean improvement to models by the inclusion of each feature in the hyperparameter optimisation of \texttt{DMO-opt-sat}. For each feature, there are 512 points in the Sobol sequence which include that feature and 512 which exclude it. We then compute the mean of $M$ for both of these 512 combinations of parameters and find the difference. Given lower values of $M$ are better, we show the differences such that large negative values correspond to substantial average improvement to the models. We exclude any points where $M=100$ due to the model exceeding our maximum memory cap when computing the means. Note the scale in the y-axis due to $M_{\rm{DM, 10kpc}}$ having such a large impact on the models compared with any other feature.}
        \label{fig:sobolbar}
    \end{figure}
\end{center}
\section{Appendix A: Principal Component Analysis (PCA) for the Input Features}
\label{ap:pca}

An investigation into the principal components of the input features was performed to see if there was opportunity for dimensionality reduction. This would allow for a very similar amount of information about the dark matter halos to be encoded in a smaller number of features, improving the computational efficiency of the stellar mass predictive model trained on these features.

Figure \ref{fig:pca} shows the cumulative fractional variance as a function of the number of components used in the PCA, looking at data in the training set from L1\_m8. In a situation where dimensionality reduction would be productive, we'd see a sharp increase in the variance in small nubers of components. However, for this set of input features we see a gradual increase in the cumulative variance, indicating that there would be substantial information loss by using only a small number of components. Therefore, the idea of using PCA for dimensionality reduction was discarded.

\section{Appendix B: Training and predicting using the HYDRO-opt model}
\label{ap:smhm}

For the purpose of comparison with \texttt{DMO-opt}, we construct a pair of models trained solely on data from L1\_m8 and trained on all features from that simulation. We refer to this pair of models as \texttt{HYDRO-opt}. We refer to the centrals model as \texttt{HYDRO-opt-cent} and the satellites model as \texttt{HYDRO-opt-sat}. We also optimised the hyperparameters of these models using a Gaussian Processes based emulator with 128 initial points selected using a Sobol sequence. We evaluated a further 965 points for \texttt{HYDRO-opt-cent} and 1172 further points for \texttt{HYDRO-opt-sat} before selecting the optimal hyperparameters. The optimal hyperparameters chosen are given in Table \ref{table:hyperparamchoices}. The validation set used for this process was the 10\% of the training set from L1\_m8, so the \texttt{HYDRO-opt} pipeline has not been applied to any DMO data as part of the hyperparameter optimisation or training. We also train a pair of regressor models with the same hyperparameters as \texttt{HYDRO-opt-cent} and \texttt{HYDRO-opt-sat} for comparison.

In Figure \ref{fig:smf-p1} we show the SMF from L1\_m8 test set as well as the predictions from \texttt{HYDRO-opt} and the regressor model with the same hyperparameters when applied to that test set. Given the similarity between the training and test sets used for these models, we see strong agreement with residuals well below the 0.2\;dex threshold over the entire calibrated stellar mass range where the SMF is reasonable.

In Figure \ref{fig:smhmfull} we show the full 2D distributions of galaxies with respect to their stellar mass and dark matter mass. This is shown for the true FLAMINGO L1\_m8 galaxies as well as the predictions made from \texttt{HYDRO-opt} and from the regressor model. We see that both models overpredict towards the modal regions of the distributions and underpredict around the tails. However we see this effect is far more pronounced for the regressor model compared with \texttt{HYDRO-opt} as we see higher fractional error in the edges of the distribution for the regressor, especially in the region $M_{\rm{DM}} > 10^{12}\rm{M_{\odot}}$. In these regions, the error number density for the \texttt{HYDRO-opt} predictions is around $-0.5$ whereas it is around $-1.5$ for the regressor predictions. This demonstrates the improvement to the characterisation of the scatter in stellar masses by our classification method.

\section{Appendix C: Additional Information on Hyperparameter Optimisation}
\label{ap:hyper}

For the hyperparameter optimisation, we use the \texttt{scikit-optimize} Gaussian Processes emulator to minimise the quantity $\mathcal{F}$ defined in Equation \ref{eq:metrmain}. $\mathcal{F}$ contains six terms added together linearly. The first three ($\Delta(16), \Delta(50),\Delta(84)$) measure the model's ability to reproduce the percentiles in the stellar mass-halo mass relation. The last three ($Q(16), Q(50),Q(84)$) give a measure of the quantisation error produced by the model having a finite number of trees and a finite number of stellar mass bins. The finite number of trees $N_{\rm{trees}}$ gives rise to quantisation errors as the probability assigned to each stellar mass bin corresponds to the fraction of trees which predict that bin. Therefore the probability must be a positive integer multiple of $1/N_{\rm{trees}}$.

In Figure \ref{fig:quanterror}, we show $Q(16)$, $Q(50)$ and $Q(84)$ in the panels from left to right as functions of the number of stellar mass bins and trees. We find that increasing the number of trees almost always reduces the quantisation error. However, to not incur a large error in the median, it is important to have a greater number of trees than the number of bins.

For the \texttt{DMO-opt-cent} optimisation, we use 128 initial random points. For optimising the \texttt{DMO-opt-sat}, we use a larger Sobol sequence with 1024 random points to initialise the emulator. The choice to use a larger number of initial points is due to the expanded parameter space which now includes the flags on each dark matter feature for whether that feature should be used. 

In Figure \ref{fig:sobolscatter}, we show the distribution of $\mathcal{F}$ for the 1024 Sobol sequence points from the optimisation of the \texttt{DMO-opt-sat} for each hyperparameter. We find increasing the number of trees increases the performance of the models, though with diminishing improvements as the number increases. We also find that increasing the number of stellar mass bins increases the performance of the model with diminishing returns also. On the other hand, there are less noticeable correlations between $\mathcal{F}$ and the maximum depth of trees, the number of features to use per tree and the minimum samples at which to split. We find unsurprisingly that increasing the fraction of the data which is used for training strongly correlates with the size of the trained model. Furthermore we find that increasing the amount of data trained on strongly correlates with the size of the trained model.

In Figure \ref{fig:sobolbar}, we show how the inclusion/exclusion of each dark matter feature affects the performance of our \texttt{DMO-opt-sat} models. As a Sobol sequence samples evenly across the parameter space, there are 512 samples which include each feature in training and 512 samples which do not. We therefore compute the mean of $\mathcal{F}$ for all models which include each feature and the mean for all that do not, which we show the difference of in Figure \ref{fig:sobolbar}. We find that including $\sigma_{\rm{DM}}$ gives the largest average improvement to our models, as it correlates strongly with $M_{\star}$. On the other hand, we find $M_{\rm{DM, 10kpc}}$ to have the most negative impact on our model. This is what we expect based on Figure \ref{fig:meandiffsdmo} as it has substantial difference in its distribution between the hydrodynamic and DMO runs. $R_{\rm{max}}$ also shows a similar discrepancy in Figure \ref{fig:meandiffsdmo}, but we do not see this reflected in Figure \ref{fig:sobolbar} as the correlation of $R_{\rm{max}}$ with $M_{\star}$ is lower than the correlation of $M_{\rm{DM, 10kpc}}$ with $M_{\star}$ so models are less likely to select $R_{\rm{max}}$ as a feature to split on. Despite the high improvement to $\mathcal{F}$ by including $\sigma_{\rm{DM}}$, the final \texttt{DMO-opt-sat} model does not include it. This is because including $\sigma_{\rm{DM}}$ in a model causes the model to use counterproductive features such as $M_{\rm{DM, 10kpc}}$ less frequently in its predictions due to the high correlation of $\sigma_{\rm{DM}}$ with $M_{\star}$, thus improving the quality of predictions only in the cases where highly detrimental features are also used.

\end{appendix}

\end{document}